\documentclass[twocolumn]{aastex631}

\usepackage{float}
\usepackage{amsmath}

\submitjournal{ApJ}

\begin{document}

\title{
Investigating Turbulence Effects on Magnetic Reconnection Rates Through Three-Dimensional Resistive Magnetohydrodynamical Simulations
}

\author[0000-0002-7009-9232]{Giovani H. Vicentin}
\affiliation{Departamento de Astronomia, Universidade de São Paulo,
Rua do Matão 1226, 05508-090, São Paulo, Brazil}

\author[0000-0002-0176-9909]{Grzegorz Kowal}
\affiliation{Escola de Artes, Ciências e Humanidades, 
Universidade de São Paulo, Rua Arlindo Bettio 1000, 03828-000, São Paulo, Brazil}

\author[0000-0001-8058-4752]{Elisabete M. de Gouveia Dal Pino}
\affiliation{Departamento de Astronomia, Universidade de São Paulo,
Rua do Matão 1226, 05508-090, São Paulo, Brazil}

\author[0000-0002-7336-6674]{Alex Lazarian}
\affiliation{Department of Astronomy, University of Wisconsin,
475 North Charter Street, Madison, Wisconsin 53706, USA}



\begin{abstract}
We investigate the impact of turbulence on magnetic reconnection through high-resolution 3D magnetohydrodynamical (MHD) simulations, spanning Lundquist numbers from $S=10^3$ to $10^6$. Building on Lazarian and Vishniac's (1999) theory, which asserts reconnection rate independence from Ohmic resistivity, we introduce small-scale perturbations until $t=0.1\, t_A$. Even after the perturbations cease, turbulence persists, resulting in sustained high reconnection rates of $V_\text{rec}/V_A \sim 0.03-0.08$. These rates exceed those generated by resistive tearing modes (plasmoid chain) in 2D and 3D MHD simulations by factors of 5 to 6. Our findings match observations in solar phenomena and previous 3D MHD global simulations of solar flares, accretion flows, and relativistic jets. The simulations show a steady-state fast reconnection rate compatible with the full development of turbulence in the system, demonstrating the robustness of the process in turbulent environments. We confirm reconnection rate independence from the Lundquist number, supporting Lazarian and Vishniac's theory of fast turbulent reconnection. Additionally, we find a mild dependence of $V_\text{rec}$ on the plasma-$\beta$ parameter, decreasing from 0.036 to 0.028 (in Alfvén units) as $\beta$ increases from 2.0 to 64.0 for simulations with a Lundquist number of $10^5$. Lastly, we explore the magnetic Prandtl number's ($\text{Pr}_m=\nu/\eta$) influence on the reconnection rate and find it negligible during the turbulent regime across the range tested, from $\text{Pr}_m=1$ to $60$. 

\end{abstract}



\section{Introduction} \label{sec:intro}
Magnetic reconnection, a fundamental process in plasma physics and astrophysics, plays a crucial role in converting magnetic energy into kinetic and thermal energy in the plasma. The concept of magnetic reconnection and its first model was introduced by \cite{Sweet1958} and \cite{Parker1957}, who laid the foundation for understanding this phenomenon, especially for solar flares. In their seminal works, Sweet and Parker proposed a theoretical model describing how magnetic field lines from oppositely directed magnetic fields could merge and annihilate in a current sheet, releasing vast amounts of energy in the process, which became a cornerstone for studying various astrophysical phenomena, such as solar flares \citep{Parker1957, gold1960origin, piddington1974alfven, forbes1991magnetic}, magnetospheric substorms \citep{dungey1961interplanetary, paschmann1979plasma, birn1980computer}, stellar coronal heating \citep{parker1983magnetic, parker1988nanoflares, heyvaerts1984coronal}, and a possible mechanism for the acceleration of particles \citep{speiser1965particle,sakai1988particle}. 

According to this model, usually referred to as Sweet–Parker mechanism, the reconnection takes place along the entire plane (the current sheet). The problem is that the Sweet-Parker reconnection rate is inversely proportional to the square root of the Lundquist number $S = LV_A/\eta$, where $L$ is the typical length of the system, $V_A$ is the Alfvén velocity and $\eta$ is the Ohmic resistivity. Since in astrophysical environments, we can have Lundquist numbers as high as $10^{18-20}$, the reconnection speed in this model, $V_{\text{rec}} \propto S^{-1/2}$, is negligible and cannot explain observations.

The understanding of magnetic reconnection was further refined by \cite{petschek1964physics}, who introduced a model that includes a stationary wave within the reconnection process, able to provide a short diffusion region. This model, also called X-point reconnection, enhances the reconnection rate and provides a maximum speed dependent on the Lundquist number as $V_{\text{rec}} \propto (\ln S)^{-1}$. On the other hand, the X-point model of reconnection seems to be unstable in the resistive magnetohydrodynamical (MHD) case for small values of the Ohmic diffusivity, as shown in the MHD simulations performed by \cite{biskamp1996magnetic}, where the X-point magnetic field tends to relax to a Sweet–Parker configuration, which again provides slow reconnection rates.

Resistive instabilities can occur in a magnetized fluid, producing a long-wave \textit{tearing} mode in the current sheet, which causes the magnetic field to reconnect and form a series of long narrow magnetic islands in a two-dimensional (2D) flow, as demonstrated by \cite{furth1963finite}. From this concept of breaking the current sheet into several magnetic islands, \cite{heyvaerts1984coronal} stated that tearing mode may be the cause of the solar coronal heating, while \cite{lee1985theory, lee1986multiple} proposed a model of multiple X reconnection lines  in the Earth's magnetopause. 

If the current sheet is thin enough and the tearing instability develops, a plasmoid chain (secondary islands) can be formed \citep{loureiro2007instability} at least in 2D space. Then, the plasmoids interact, merge, grow, and are advected with the 2D ﬂow. 2D MHD simulations \citep{bhattacharjee2009fast, huang2010scaling} evidenced that for high Lundquist numbers, of $S \gtrsim 4 \times 10^4$, resistive reconnection may become plasmoid dominated and evolve independently of the Ohmic resistivity $\eta$ (fast reconnection), at a universal rate of $V_\text{rec} \sim 0.01 \, V_A$. On the other hand, it is important  to underscore that these models are confined to 2D  space geometry. 


For three-dimensional (3D) MHD flows, \cite{lazarian1999reconnection} solved the problem of slow magnetic reconnection using turbulence, which is ubiquitous in astrophysics  \cite[e.g., ][]{Armstrong1995ApJ...443..209A, Chepurnov_2010}. In this scenario, the wandering of the magnetic field lines creates numerous reconnection sites that can take place simultaneously, rendering the reconnection rate explicitly independent of the Ohmic resistivity and the Lundquist number (fast reconnection). It depends only on the injection scale ($\ell$) and the injection velocity ($v_\ell$) of the turbulence, according to

\begin{equation}
    \frac{V_{\text{rec}}}{V_A} = \text{min} \left[ \left( \frac{L}{\ell} \right)^{1/2}, \left( \frac{\ell}{L} \right)^{1/2} \right] \left( \frac{v_\ell}{V_A} \right)^{2}. 
\end{equation}

Given that the velocity of the turbulence at the injection scale can be  comparable to the Alfvén speed ($v_\ell \sim V_A$) and the injection scale of turbulence can be of the same order of the typical length of the system ($\ell \sim L$), according to the \citet{lazarian1999reconnection} theory the reconnection rate may reach $V_\text{rec} \lesssim V_A$.

This theory of turbulent reconnection was successfully tested in classical \citep{kowal2009numerical, kowal2012visc, Kowal_2011, Kowal_2012, kowal2017statistics, delValle_2016, Beresnyak_2017, Kadowaki_2018}  
and relativistic \citep{takamoto2015turbulent,singh2016spatial, Takamoto_2016, Kadowaki_2021, Medina-Torrejón_2021, Medina-Torrejón_2023}, 
MHD simulations \citep[see also][for recent reviews]{Eyink_2011,Lazarian2019review,Lazarian2020review}. Particularly, \cite{kowal2009numerical, kowal2012visc}, 
obtained maximum reconnection rates of $V_\text{rec} \gtrsim 0.1 \, V_A$ - one order of magnitude higher than the ones from 2D MHD simulations of plasmoid chains. Moreover, by applying an algorithm for the search of magnetic reconnection sites in 3D MHD shearing-box numerical simulations of accretion disks, \cite{Kadowaki_2018} found even higher reconnection rates, with peak values of $V_\text{rec}/V_A  \sim 0.2$, and without evidence of plasmoid formation.

Although the \cite{lazarian1999reconnection} reconnection model was originally formulated based on three-dimensional MHD turbulence \citep{GS_1995}, it is expected to work in two-dimensional systems. However, there is currently no established theory for turbulence-induced reconnection in two dimensions, which makes it difficult to predict whether the reconnection rate will differ from that in three dimensions. Two-dimensional numerical studies of reconnection in the presence of driven turbulence have shown an increase in the reconnection rate when turbulence is present. However, its relation with the injection scale or power does not align with the predictions made by \cite{lazarian1999reconnection}. 
Notably, and in contrast to \cite{lazarian1999reconnection} theory, the reconnection rate in two dimensions depends on magnetic resistivity, as reported by \cite{Kulpa-Dybel-2010}.

The differing impact of fluctuations and turbulence on magnetic reconnection in 2D and 3D stems from the dimensionality of field line relaxation. In 3D, field line tension can be relieved through motion along the third dimension, allowing for more complex topological changes and potentially slower reconnection rates in certain scenarios. Conversely, in 2D, field line tension is confined to the plane, leading to a direct enhancement of current density and an accelerated reconnection rate in the presence of fluctuations, as there is no out-of-plane relaxation. This distinction fundamentally alters the reconnection dynamics in turbulent environments, leading to different statistical properties of energy dissipation and current sheet evolution between 2D and 3D cases. In this work, we will focus on the impact of turbulence on the reconnection rate in 3D MHD simulations of current sheets.

Turbulence is naturally induced in astrophysical environments by different instabilities, like the Kelvin-Helmholtz instability \citep{chandrasekhar1968hydrodynamic} due to sheared velocities \citep[e.g.][]{Kowal_2020} , the magnetorotational instability \citep[MRI,][]{balbus1991powerful, balbus1998instability} in accretion flows \citep[e.g.][]{Kadowaki_2018}, and {the current-driven}  kink instability \citep{Begelman_1998} in toroidal magnetic field in jets \citep[e.g.][]{singh2016spatial, Kadowaki_2021, Medina-Torrejón_2021, Medina-Torrejón_2023, dalpino2024}. Turbulence may also be self-generated by reconnection, as first discussed by \cite{lazarian1999reconnection}, and numerically tested by \cite{daughton2011role, Oishi_2015, Huang_2016, Beresnyak_2017, kowal2017statistics, Kowal_2020, beg2022evolution}. 

In particular, in their 3D MHD simulations, \cite{kowal2017statistics} obtained a weak dependence of the reconnection rate ($V_\text{rec}$) on the plasma-$\beta$, which gives the ratio of thermal to magnetic pressure. 
In their case, an initial velocity perturbation is inserted in the system, and the numerical resistivity is estimated to be $\eta \lesssim 3 \times 10^{-4}$, resulting in Lundquist numbers of $S \gtrsim 3000$. 

In this study, we expand upon these previous investigations by exploring the relationship between the reconnection rate and $\beta$ for high Lundquist numbers, reaching $S=10^5$. 
Similar to \cite{kowal2017statistics}, we observe a weak dependence of $V_\text{rec}$ on $\beta$, and the prevalence of the turbulence after the initial injection ceases. 

Additionally, we explore the influence of the magnetic Prandtl number $\text{Pr}_m = \nu / \eta$, representing the ratio of the fluid  viscosity ($\nu$) to the resistivity ($\eta$), and find this influence to be negligible across the range tested, from $\text{Pr}_m=1$ to $60$.

In addition to examining the correlation between $V_\text{rec}$ and plasma-$\beta$ and $\text{Pr}_m$, we compare our findings on turbulent reconnection with those from scenarios where fast reconnection arises solely from tearing mode instability. 
In contrast to the case of continuously forced turbulence \citep[e.g.,][]{kowal2009numerical, Loureiro2009}, one of the key objectives of this work is to investigate the self-generation and self-sustenance of turbulence in the 
current sheet without continuous forcing in 3D. For that, we injected small-scale, multi-mode perturbations peaked around a wavenumber $k = 128$ during a very short initial time interval ($\Delta t = 0.1 \, t_A$).
We observe that the perturbation scale of $\ell \sim 128^{-1}$ is sufficiently close to the dissipative/resistive scale of the simulation, making it capable of inducing instabilities at sufficiently small scales, such as tearing-mode instability. We perturb the tearing instability with oblique modes for which $ |\mathbf{k}| = k \simeq 128$, 
but $k_x < (2 \pi \delta)^{-1}$, where $\delta$ is the initial thickness of the current sheet. As the instability develops, it leads to turbulence in the system, which can be measured through the power spectrum. Additionally, other instabilities, such as the kink and the Kelvin-Helmholtz instabilities, may also arise and contribute to turbulence in the current layer.
We find that the fastest reconnection rate is achieved at the same time that turbulence is generated and sustained in the system, reaching peaks one order of magnitude higher than the tearing
regime and average rates five to six times greater than the ones generated solely by tearing-mode/plasmoid instability in 2D (see Figure \ref{fig:vrec_plasmoid_turb}).

In Section \ref{sec:numerics} we describe the setup of our simulations, the governing equations, and the numerical schemes we have used to solve them. In Section \ref{sec:recrate} we go into detail on the measurement of the reconnection rate from the time derivative of the magnetic flux. In Section \ref{sec:results} we present the results of our numerical simulations, 
mainly about the prevalence of turbulence in the system and the dependence of the reconnection rate on the plasma$-\beta$ and the magnetic Prandtl number. In Section \ref{sec:discussion} we discuss our results and in Section \ref{sec:conclusions} we set our main conclusions.

\section{Numerical Setup} \label{sec:numerics}

We use the high-order shock-capturing Godunov-type code AMUN\footnote{The code is freely available at \url{https://bitbucket.org/amunteam/amun-code/}.} \citep{kowal2009numerical, kowal2012visc} to solve the isothermal visco-resistive 3D MHD equations:

\begin{equation}
    \frac{\partial \rho}{\partial t} + \nabla \cdot (\rho \mathbf{v}) = 0,
\end{equation}

\begin{equation}
    \frac{\partial (\rho \mathbf{v})}{\partial t} + \nabla \cdot \left[ \rho \mathbf{v}\mathbf{v} + \left( p + \frac{B^2}{8\pi} \right) \mathbf{I} - \frac{1}{4\pi} \mathbf{B} \mathbf{B} \right] = \nabla \cdot \mathbf{\tau} + \mathbf{f},
\end{equation}

\begin{equation}
    \frac{\partial \mathbf{B}}{\partial t} + \nabla \times \mathbf{E} = 0,
\end{equation}

\noindent
where $\rho$ and $\mathbf{v}$ are the plasma density and velocity, respectively, $\mathbf{B}$ is the magnetic field, $\mathbf{E} = - \mathbf{v} \times \mathbf{B} + \eta \mathbf{j}$ is the electric field, $\mathbf{j} = \nabla \times \mathbf{B}$ is the current density, $p = \rho c_s^2$ is the thermal pressure, $c_s$ is the sound speed, $\eta$ is the resistivity coefficient, $\mathbf{\tau} = \nu \rho \left[ \nabla \mathbf{v} + \nabla^T \mathbf{v} - \frac{2}{3} \nabla \cdot \mathbf{v} \right]$ is the viscous stress tensor, $\nu$ is the kinematic viscosity, and $\mathbf{f}$ represents the forcing term. 

To numerically solve the 3D MHD set of equations, we used the HLLD Riemann solver \citep{HLLD} with a $5^\mathrm{th}$-order Monotonicity-Preserving (MP5) reconstruction method \citep{mp5} to reconstruct the Riemann states, and a $3^\mathrm{rd}$-order 4-step Embedded Strong Stability Preserving Runge-Kutta (SSPRK) method for time advance, where the time step is controlled by both the Courant–Friedrichs–Lewy (CFL) condition and the integration error \citep[see, e.g.,][]{SSPRK324}.

The code uses dimensionless equations in such a way that the strength of the magnetic field is expressed in terms of the Alfvén velocity, which is defined by the antiparallel component of the magnetic field (reconnecting field) and the unperturbed density $\rho_0 = 1$. All other velocities are expressed as fractions of the Alfvén speed, the length of the box in the $x-$direction ($L_x$) defines the unit of distance, and time is measured in units of the Alfvén time, defined as $t_A \equiv L_x/V_A$.

\begin{figure}[t]
    \centering
    \includegraphics[width = 0.49 \textwidth]{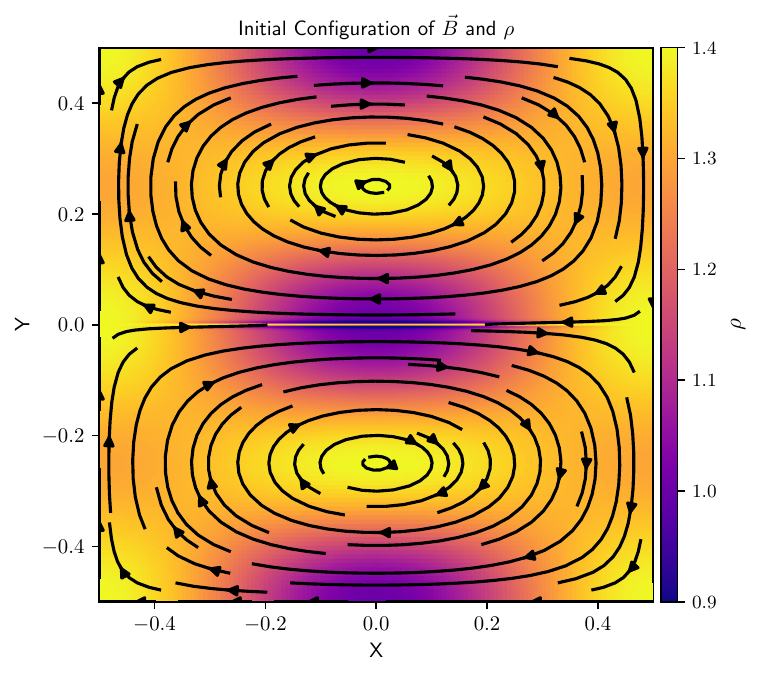}
    \caption{Initial configuration of magnetic field and density. Black arrows represent the in-plane component of the magnetic field, and the colormap is the density profile. The out-of-plane component of the magnetic field $(B_z)$ is set to be constant.}
    \label{fig:initial-config}
\end{figure}

The configuration of the magnetic field in the reconnection region is similar to the one employed by \cite{bhattacharjee2009fast, Huang_2016, beg2022evolution}, where the attraction between two coalescing magnetic flux tubes is the driver of magnetic reconnection. The simulation domain is a 3D box with dimensions $L_x = L_y = 1$ and $L_z = 0.5$, centered at $(0,0,0)$, and the effective resolution for all the simulations was set to be $1024 \times 1024 \times 512$ in a uniform grid, in such a way that each grid cell has size $\Delta x = \Delta y = \Delta z = h = 1/1024$. 
The reconnecting magnetic field is  along the $x-$direction (see Fig. \ref{fig:initial-config}). 
The boundary conditions in our models are perfectly conducting, free slipping boundaries along $x$ and $y$ directions, and periodic along $z$ direction.

The initial magnetic field is given by $\mathbf{B} = \hat{z} \times \nabla \psi + B_z \, \hat{z}$, where 

\begin{equation}
    \psi = \frac{1}{2 \pi} \tanh \left(\frac{y}{\delta} \right) \cos (\pi x) \sin (2\pi y), \label{eq:psi_HB}
\end{equation}

\noindent
and $B_z$ is the guide field. We adopted, for our initial configuration, a constant guide field, while the density is non-uniform in order to maintain the pressure balance (see Fig. \ref{fig:initial-config}). The initial current sheet thickness $\delta$ is set to be $S^{-1/2}$, following the (laminar) Sweet-Parker relation.

In most simulations analyzed in this work, 
we used the standard Lundquist number of $S=10^5$. There, the thickness of the laminar Sweet-Parker regime of $\delta \sim 0.0032$ is well-resolved at the grid size of $h=1/1024$. Moreover, we observe that in our 3D turbulent simulations, turbulence increases the average thickness of the current sheet, reaching average values of $\langle \delta \rangle = 0.0070(9)$ for the fully turbulent regime. {In this case, the current sheet's thickness spans more than seven grid cells}. To verify whether the current sheet thickness is well-resolved in our simulations, we have also conducted a convergence analysis using different resolutions for simulations with $S=10^5$. Our results showed that in our 3D simulations, the grid sizes of $h=1/1024$ and $h=1/2048$ yield similar reconnection rates {(see Fig. \ref{fig:compar_resolutions_1k_2k} and the discussion in Appendix \ref{appendix:convergence_pinj}).} 

For comparison, we have also considered a set of models with no initial external perturbation injection. These models were tested for Lundquist numbers between $S=10^3 - 10^6$.  

In the initial configuration,  to stabilize the flux ropes against the kink instability, 
we implemented a guide field with amplitude $B_z \ge 0.5$ for all models simulated.

The stability of a cylindrical magnetic flux rope against kink is often quantified in terms of the Kruskal-Shafranov factor \citep{shafranov1956stability, kruskal1958instability}, given by

\begin{equation}
    \zeta = \dfrac{2\pi R}{L_z} \dfrac{B_z}{B_\phi}, \label{eq:KSfactor}
\end{equation}

\noindent
where $R$ is the radius of the flux tube, $L_z$ is the length 
in the periodic ($z$) direction, which sets the wavelength of the longest mode, and $B_\phi$ and $B_z$ are the poloidal and toroidal fields, respectively. If $\zeta > 1$, the kink is stabilized. In our case, this criterion is satisfied by using a guide field with an amplitude of $B_z \ge 0.5$. We note, however, that the kink instability may develop during the evolution of the system, as observed in the simulation reported in \cite{beg2022evolution}.

We also tested the initial configuration as implemented by \cite{Huang_2016}, where the initial density is constant and the guide field $B_z$ is non-uniform such that the system is approximately force-balanced, and we obtained the same reconnection rates (see Appendix \ref{appendix:dif_initial_cond}).

\subsection{Initial perturbation } \label{subsec:turbulence}

As remarked, one of the main goals of this work is to study the dependence of the reconnection rate on the plasma$-\beta$ and  the magnetic Prandtl number for turbulent reconnection, as well as the prevalence of turbulence in the simulated domain. In the first attempt to numerically test the model of turbulent reconnection proposed by \cite{lazarian1999reconnection}, \cite{kowal2009numerical, kowal2012visc} performed 3D MHD simulations of current sheets where forced turbulence is continuously injected into the domain. 
Here, instead, we drive small-scale 
perturbation 
for a short initial period into the system and examine the properties and effects of the instability and the turbulence that follows.


In order to introduce the 
perturbation in the system, we employ 
a technique outlined by \cite{alvelius1999random} \citep[see also][for a detailed discussion of this method]{kowal2009numerical, kowal2012visc}. This forcing is applied in spectral space, concentrated around a wave vector $k_{\text{inj}}$ that corresponds to the injection scale $l_{\text{inj}} \sim k_\text{inj}^{-1}$. Within a shell extending from $ k_\text{inj} - \Delta  k_\text{inj}$ to $ k_\text{inj} + \Delta  k_\text{inj}$, we disturb $N$ discrete Fourier components of velocity using a Gaussian profile characterized by a half-width $k_c$ and a peak amplitude $v_f$ at the injection scale. The amplitude of the turbulence is given by the injection power $P_\text{inj}$.

We consider $k_\text{inj} = 128$, $\Delta k_\text{inj} = 0.01$,  $N = 947$ driving modes, and the   
perturbation is injected in the system at the beginning of the simulation ($t=0$) up to $t=0.1 \, t_A$.
We adopted two different injection powers, of $P_\text{inj} = 0.1$ and $0.5$.

\begin{table}[t]
    \centering
\begin{tabular}{cccccccc}
$\eta (10^{-5})$ & $\text{Pr}_m = \frac{\nu}{\eta}$  & $\beta = \frac{p_\text{th}}{p_\text{mag}}$ & $P_{\text{inj}}$ & $t_\text{max}$ \\ \hline \hline
100.0 & 1.0 & 0.5 & 0.0 & 3.0  \\
100.0 & 1.0 & 2.0 &  0.0 & 3.0 \\
100.0 & 1.0 & 8.0 &  0.0 & 3.0 \\
100.0 & 1.0 & 18.0 &  0.0 & 3.0 \\
100.0 & 1.0 & 32.0 &  0.0 & 3.0 \\ \hline
10.0 & 1.0 & 0.5 &  0.0 & 3.0 \\
10.0 & 1.0 & 2.0 &  0.0 & 3.0 \\
10.0 & 1.0 & 8.0 &  0.0 & 3.0 \\
10.0 & 1.0 & 18.0 &  0.0 & 3.0 \\
10.0 & 1.0 & 32.0 &  0.0  & 3.0 \\ \hline
1.0 & 1.0 & 2.0 &  0.0  & 3.0  \\
1.0 & 1.0 & 8.0 &  0.0  & 3.0 \\
1.0 & 1.0 & 32.0 &  0.0 & 3.0 \\ 
0.1 & 1.0 & 2.0 &  0.0  & 3.0 \\ \hline
1.0 & 1.0 & 2.0 &  0.1  & 3.0 \\
1.0 & 10.0 & 2.0 &  0.1 & 5.0  \\
1.0 & 30.0 & 2.0 & 0.1 & 5.0 \\
1.0 & 40.0 & 2.0 & 0.1 & 5.0 \\
1.0 & 50.0 & 2.0 & 0.1 & 5.0 \\
1.0 & 60.0 & 2.0 & 0.1 & 5.0 \\ \hline
1.0 & 1.0 & 2.0 & 0.5  & 3.0 \\
1.0 & 1.0 & 32.0 &  0.5 & 3.0 \\
1.0 & 1.0 & 64.0 &  0.5 & 3.0 \\ \hline
1.0 & 1.0 & 2.0 &  0.1  & 3.0 \\
1.0 & 1.0 & 32.0 &  0.1 & 3.0 \\
1.0 & 1.0 & 63.0 &  0.1 & 3.0
\end{tabular}

    \caption{List of initial parameters of the simulated models. 
    In this table, $\eta$ is the explicit resistivity, $\text{Pr}_m$ is the magnetic Prandtl number, the ratio between viscosity and resistivity, $\beta$ is the ratio between thermal and magnetic pressures, $P_\text{inj}$ is the injection power of the initial perturbation, and $t_\text{max}$ represents the time (in Alfvén units) we stop the simulation.}
    \label{tab:list_models}
\end{table}

Table \ref{tab:list_models} lists the initial conditions of the models simulated in this work. As remarked earlier, the models  without an initial perturbation have Lundquist numbers in the range $10^3 \le S \le 10^6$, while those with an initial perturbation are set to $S=10^5 $. 

\section{Measuring the reconnection rate}\label{sec:recrate}


We adapt the method described by \cite{kowal2009numerical} to calculate the reconnection rate from the unsigned magnetic flux. Specifically, we integrate $|B_x|$ over the entire $yz-$plane ($x=0$), perpendicular to the current sheet. Since the signed flux through this plane is zero, dividing the unsigned integral by two gives the flux contribution from each polarity. As time evolves, the two initial flux ropes merge, and $|B_x|$ decreases due to the reconnection process. Then, after normalizing the flux by the length of the box in the $z$ direction, we have the reconnection rate given by the time derivative of the unsigned magnetic flux:







\begin{equation}
    V_{\text{rec}} = - \frac{1}{2 |B_{x,0}| L_z} \frac{\partial}{\partial t}\iint_S |B_x| \, dy \, dz , \label{eq:vrec_kowal09_thiswork}
\end{equation}

\noindent where $|B_{x,0}|$ represents initial amplitude of the non-reconnecting field, that in this configuration is given by $|B_{x,0}| = \max (|B_x|_{x=0}) \sim 1$.

Since for all the simulations performed we adopted $L_z = 0.5$, the reconnection rate becomes simply $V_\text{rec} = - \partial_t \Phi_B$, where $\Phi_B$ is the unsigned magnetic flux calculated across the plane $x=0$. Therefore, we end up with a similar way to compute the reconnection rate as implemented by \cite{Huang_2016}, where $V_{\text{rec}}$ is calculated as the time derivative of the magnetic flux across the current sheet ($xz-$plane). On the other hand, by calculating the reconnection rate along the $x-$direction, they may have an influence on the accumulation of magnetic flux at the left and right boundaries of the box, which is prevented by using the method described above, where the magnetic flux is integrated in the plane perpendicular to the current sheet and far from the $x-$boundaries.

\section{Results} \label{sec:results}

We discuss in this Section the results of 3D MHD sim-
ulations of current sheets first in the absence and then,
in the presence of initial perturbations, for different val-
ues of plasma$-\beta$, Lundquist and Prandtl numbers.

\subsection{Models with no initial external perturbation}
\begin{figure}[t]
    \centering
    \includegraphics[width = 0.45 \textwidth]{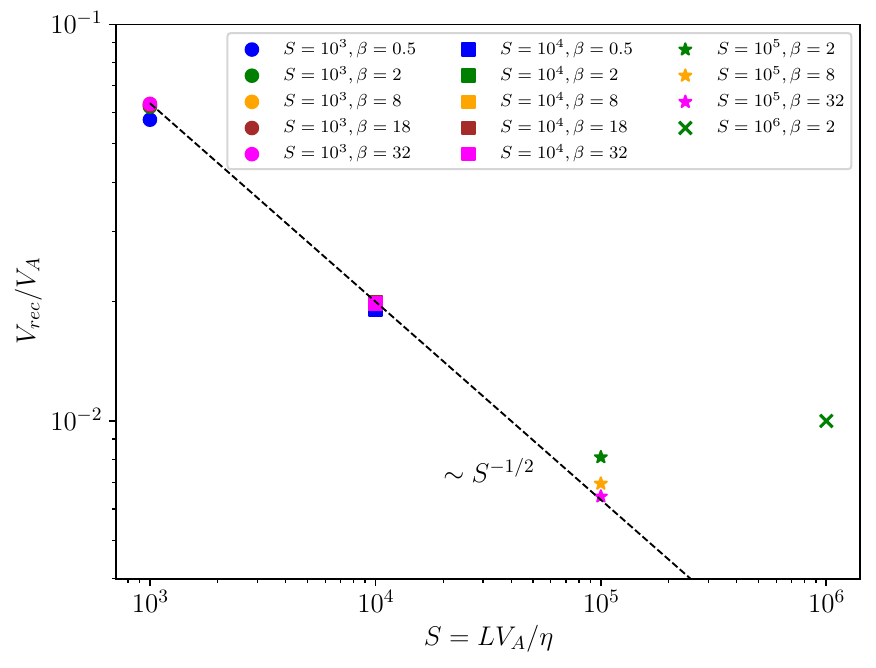}
    \caption{Dependence of the reconnection rate $V_{\text{rec}}$ on the Lundquist Number $S$ for simulations without forcing at early times, and for different values of plasma$-\beta$. The dashed black line represents the Sweet-Parker scaling of $V_{\text{rec}} \propto S^{-1/2}$. Some circles and squares are overlapped. This is the case for intermediate values of $\beta$, between 2 and 18, for both $S=10^3$ and $10^4$.
    }
    \label{fig:Vrec_depS_beta}
\end{figure}

In the absence of forced turbulence injection or instabilities, the reconnection rate measured within the current sheet should follow the Sweet-Parker dependence on the Lundquist number, $V_{\text{rec}} \sim S^{-1/2}$. Figure \ref{fig:Vrec_depS_beta} shows the reconnection rate measured using Eq. (\ref{eq:vrec_kowal09_thiswork}), for different values of Lundquist number and plasma$-\beta$ for this case, where the initial system did not suffer any perturbation.

We notice a good agreement between our data and the Sweet-Parker dependence of $V_{\text{rec}}$ on the Lundquist number (dashed black line in Fig. \ref{fig:Vrec_depS_beta}) for $S < 10^5$. For higher Lundquist numbers ($S \ge 10^5$) the system becomes unstable to the tearing mode (plasmoid) instability \citep[see, e.g.][]{loureiro2007instability, bhattacharjee2009fast} and the reconnection rate starts to deviate from the Sweet-Parker fit, $V_{\text{rec}} \sim S^{-1/2}$, becoming approximately independent on the Lundquist number and reaching maximum values of $V_{\text{rec}} \sim 0.01\, V_A$.

In Figure \ref{fig:Vrec_depS_beta}, colors represent different values of plasma$-\beta$ for a fixed magnetic field strength (or, equivalently, different sound speed). For the simulations with up to $S = 10^4$, we do not notice a substantial dependence of $V_\text{rec}$ on $\beta$, as expected for a Sweet-Parker configuration of reconnection. Contrarily, for the three simulations with $S=10^5$, where resistive instabilities may develop and drive turbulence \citep[see e.g.][]{Huang_2016, beg2022evolution}, a weak dependence of $V_\text{rec}$ on $\beta$ is observed, and higher reconnection rates are measured for lower values of $\beta$ {(or higher magnetization)}. 
The same behaviour was observed in the 3D MHD simulations of self-generated  turbulence 
 performed by \cite{kowal2017statistics}, but for lower Lundquist number, of $S \sim 3 \times 10^3$.

In the simulations conducted by \cite{kowal2017statistics}, the initial disturbance was set through a 
velocity perturbation with a random distribution of directions in the region near the 
magnetic field discontinuity. Conversely, in the set of simulations of Figure \ref{fig:Vrec_depS_beta}, where we examine the Sweet-Parker regime, no noise was introduced into the system. 
In the turbulent scenarios that we will discuss in the following sections, we initiate the simulations by introducing small-scale perturbations across the entire domain, up to $t=0.1 \, t_A$.

\subsection{Models with initial external perturbation}


We discuss in this Section the results of 3D MHD simulations of current sheets in the presence of initial multi-mode, small-scale perturbations, for models with $S=10^5$ and different values of plasma$-\beta$ and Prandtl numbers, as shown in Table \ref{tab:list_models}. As stressed, multi-mode perturbation is initially driven in the system at a high wavenumber ($k_{\text{inj}} = 128$) with the injection power of $P_{\text{inj}} = \{0.1, 0.5\}$, until $t=0.1 \, t_A$. After that, the external injection stops, and the system evolves until $ t_{\text{max}} = 3.0 \, t_A$ or $ t_{\text{max}} = 5.0 \, t_A$, depending on the model (simulations with high $\text{Pr}_m$ require more time to the turbulence become fully-developed).


\begin{figure*}[t]
    \centering
    \includegraphics[width = 0.999 \textwidth]{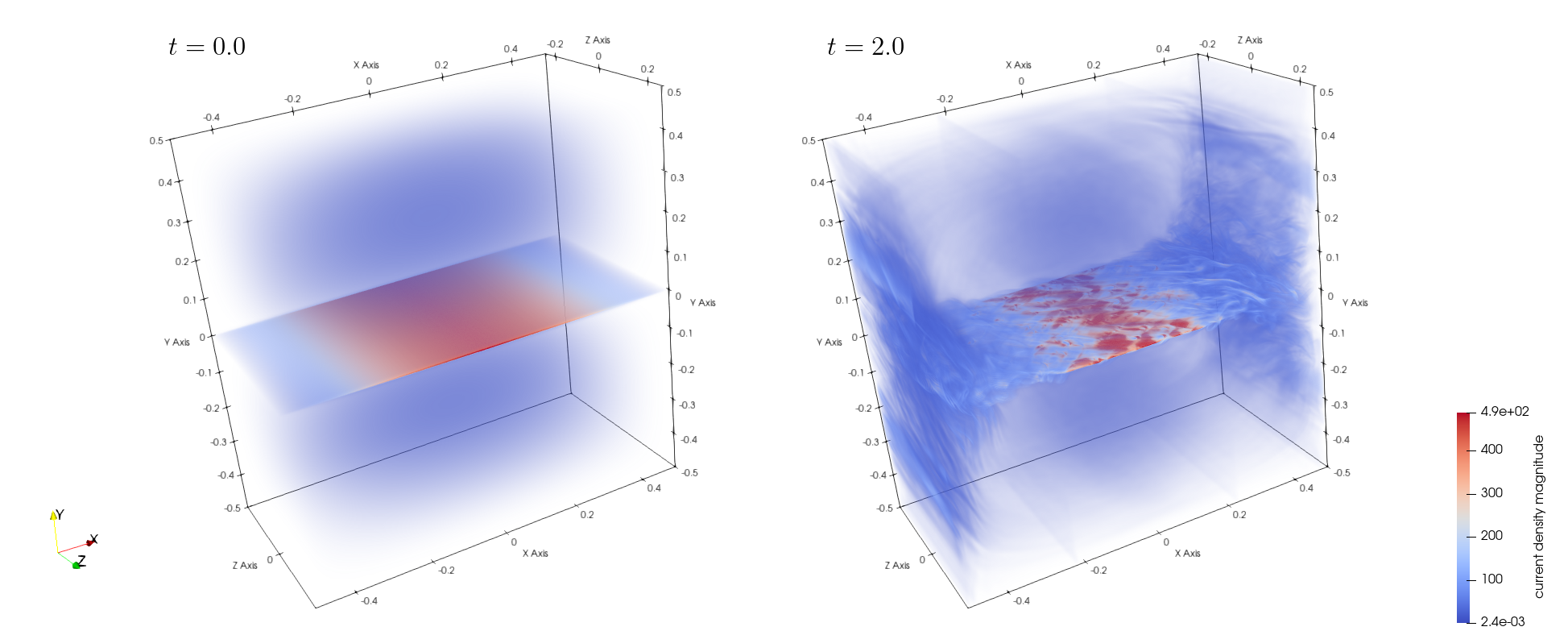}
    \caption{3D visualization of the current density magnitude, $|\mathbf{J}|$, at $t = 0$ (left) and $t=2.0$ (right) for the simulation with $S=10^5$, $\text{Pr}_m=1$, $B_z = 0.5$ and $\beta = 2.0$. Perturbation is injected up to $t=0.1 \, t_A$ with $k_\text{inj}=128$ and $P_{\text{inj}}=0.5$.}
    \label{fig:3D_visu_t0_t2}
\end{figure*}

\begin{figure*}[ht!]
    \centering
    \includegraphics[width = 0.99 \textwidth]{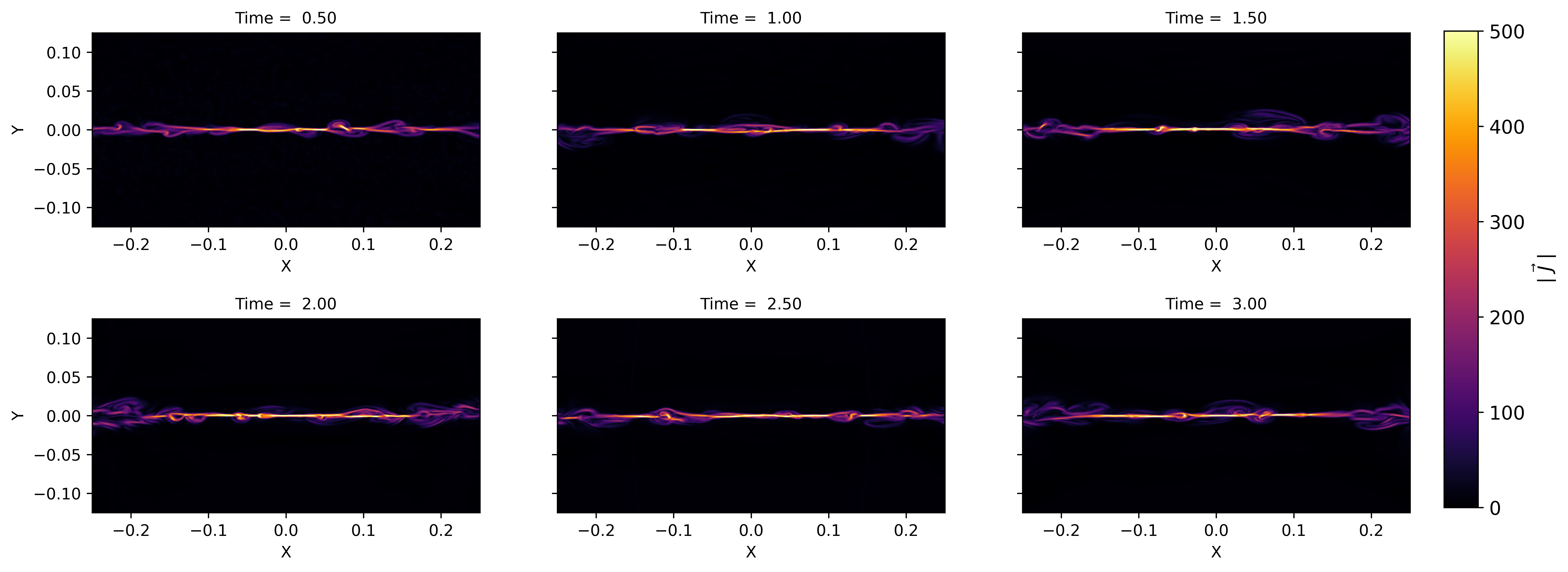}
    \includegraphics[width = 0.99 \textwidth]{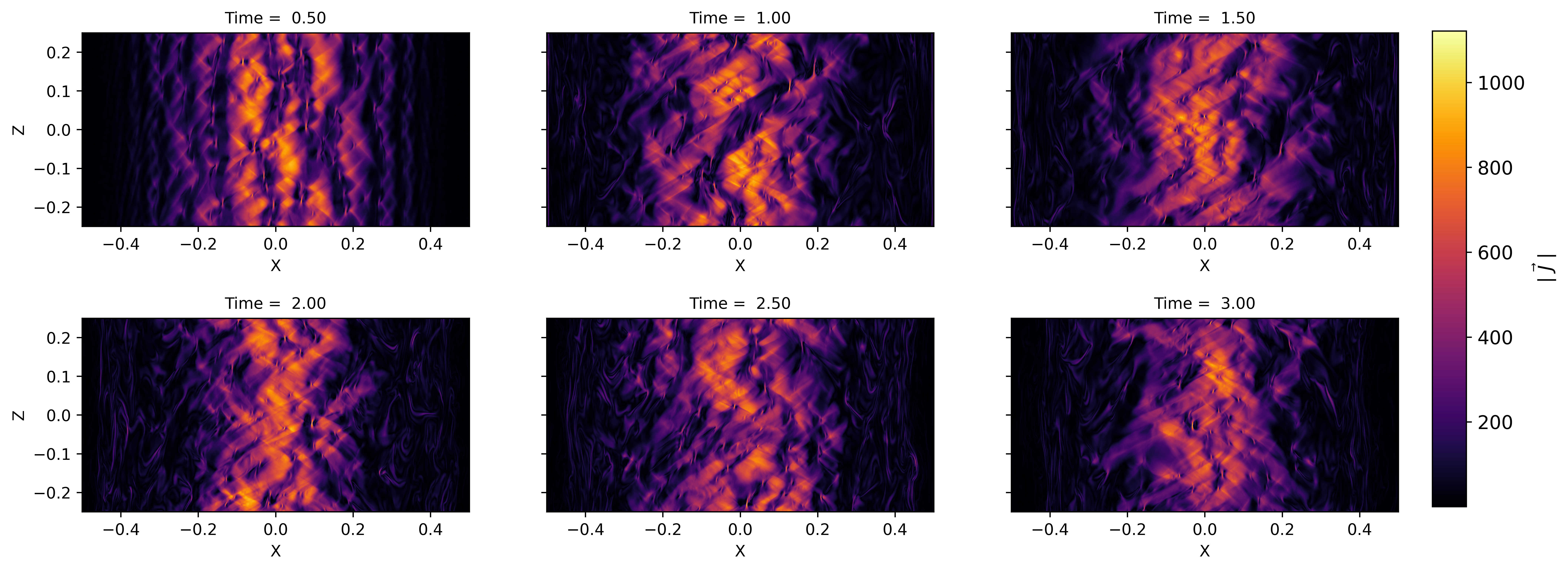}
    
    \caption{Colormaps of 2D cuts ($xy-$ and $xz-$planes) of the current density magnitude at different times for the same simulation shown in Fig. \ref{fig:3D_visu_t0_t2}. 
    }
    \label{fig:2Dcuts_currdens_S1e5_bt2_P0.5_k128}
\end{figure*}

\begin{figure*}[ht!]
    \centering
    \includegraphics[width = 0.99 \textwidth]{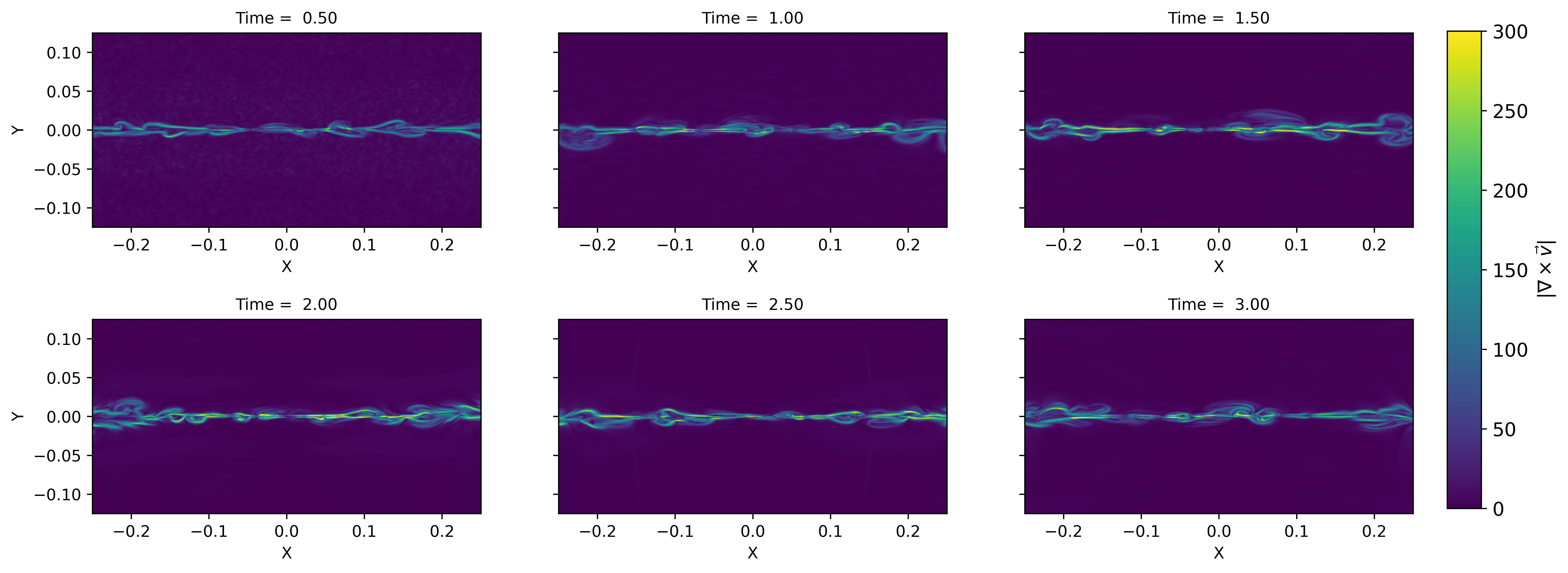}
    \includegraphics[width = 0.99 \textwidth]{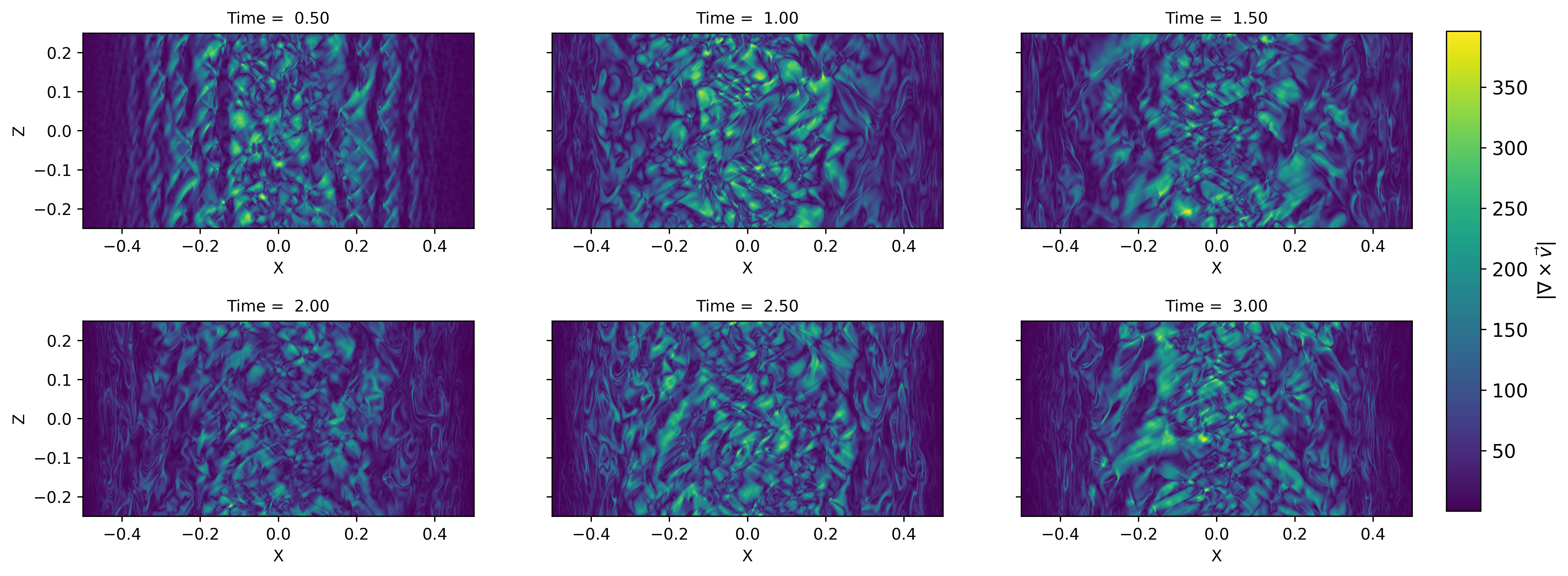}
    
    \caption{Colormaps of 2D cuts ($xy-$ and $xz-$planes) of the vorticity magnitude $(|\omega| = |\nabla \times \mathbf{v}|)$ at different times for the same simulation shown in Fig. \ref{fig:3D_visu_t0_t2}. }
    \label{fig:2Dcuts_vorticity_S1e5_bt2_P0.5_k128}
\end{figure*}



Figure \ref{fig:3D_visu_t0_t2} illustrates the 3D representation of the current density magnitude $|\mathbf{J}|$ at $t=0$ (left) and $t = 2.0 \, t_A$ (right) for the simulation with parameters 
$\text{Pr}_m=1$, $\beta = 2.0$, $B_z = 0.5$, and $P_\text{inj} = 0.5$. At $t = 2.0$, it is evident that magnetic flux accumulates along the $x-$boundaries, and turbulent structures persist in the current sheet mid-plane, even after external forcing ceases at $t = 0.1 \, t_A$. This persistence suggests the potential self-sustainability of turbulence within the simulated domain.


\begin{figure*}[t]
    \centering
    \includegraphics[width = 0.49 \textwidth]{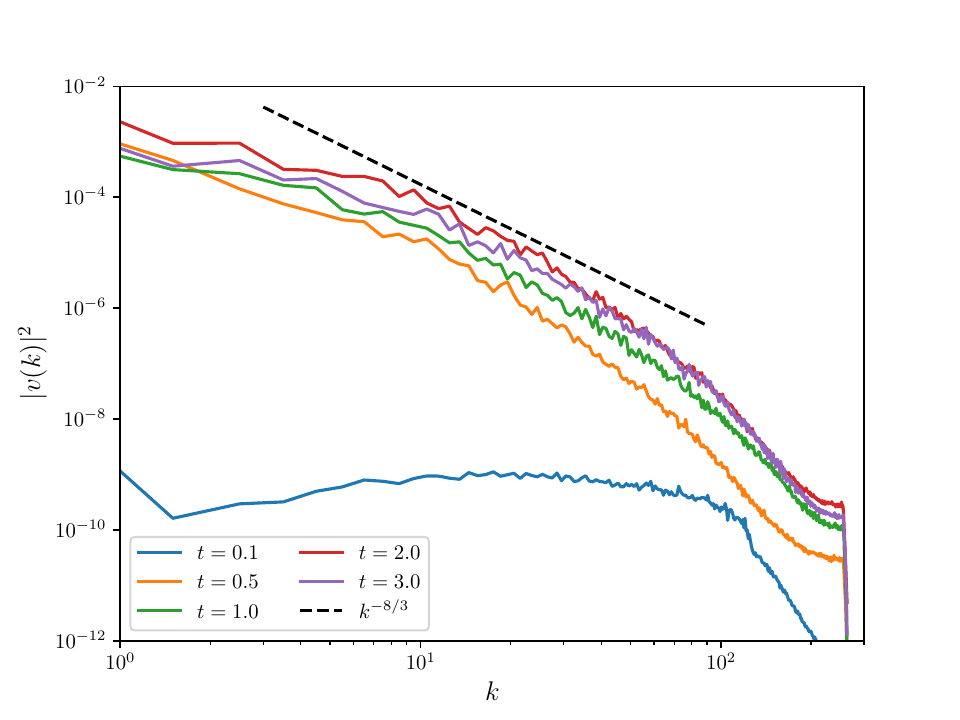} \includegraphics[width = 0.49 \textwidth]{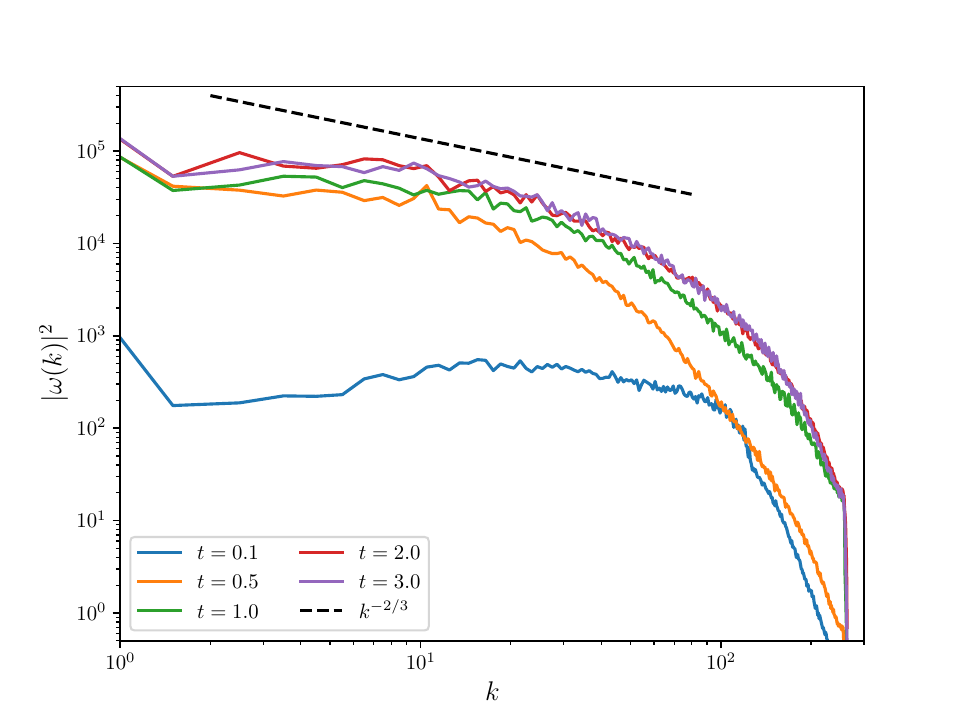}
    \caption{Power Spectra of the velocity ($\mathbf{v}$, left) and vorticity ($\omega = \nabla \times \mathbf{v}$, right) fields in the $xz-$plane (current sheet mid-plane) at different timesteps of the simulation with $S = 10^5$, $\beta=2.0$, $\text{Pr}_m = 1$, and $P_\text{inj} = 0.5$. The black dashed lines represents the 1D Kolmogorov turbulent power spectra of $\mathcal{P}(k) \propto k^{-8/3}$ (velocity) and $\mathcal{P}(k) \propto k^{-2/3}$ (vorticity).}
    \label{fig:powerspec_vorticity}
\end{figure*}

In Figures \ref{fig:2Dcuts_currdens_S1e5_bt2_P0.5_k128} and \ref{fig:2Dcuts_vorticity_S1e5_bt2_P0.5_k128}, we display the time evolution  of the current density and vorticity ($|\omega| = |\nabla\times \mathbf{v}|$) magnitudes, respectively, for the same simulation depicted in Fig. \ref{fig:3D_visu_t0_t2}. These figures present 2D cross-sections along the $xy-$ (top) and $xz-$ (bottom) planes. Once more, we observe turbulent structures persisting in the current sheet mid-plane ($xz-$plane) for extended durations. Consequently, we computed the power spectrum for the velocity and vorticity fields within this plane to ascertain the potential self-sustainability of turbulence.


To calculate the power spectrum for both the velocity and the vorticity fields, we select the data within the plane $y=0$, subtract the mean quantities (velocity and vorticity, respectively) on the mid-plane, and we perform the Fast Fourier Transform (FFT) on the input field to obtain the Fourier coefficients. Given a vector field $\mathbf{w}(\mathbf{x}) = (w_x(\mathbf{x}), w_y(\mathbf{x}), w_z(\mathbf{x}))$, the Fourier Transform of each component is:

\begin{equation}
    \tilde{w}_j(\mathbf{k}) = \mathcal{F}[w_j(\mathbf{x})] = \int w_j(\mathbf{x}) e^{-i \mathbf{k} \cdot \mathbf{x}} \, d\mathbf{x} ,
\end{equation}

\noindent where $j=\{x, y, z\}$ and $\mathbf{k} = (k_x, k_y, k_z)$ is the wave vector. Then, we calculate the integrated power spectrum $P(\mathbf{k})$ of the vector field, which is given by:

\begin{equation}
    P(\mathbf{k}) = \sum_j  |\tilde{w}_j(\mathbf{k})|^2 = \sum_j  \tilde{w}_j(\mathbf{k}) \tilde{w}_j^*(\mathbf{k}),
\end{equation}

\noindent where $\tilde{w}_j^*(\mathbf{k})$ is the complex conjugate of $\tilde{w}_j(\mathbf{k})$.

Finally, we bin the power spectrum over spherical shells in wavevector space by computing:

\begin{equation}
    P(k) = \sum_{\mathbf{k}, |\mathbf{k}| = k} P(\mathbf{k}),
\end{equation}

\noindent where the sum runs over all discrete Fourier modes $\mathbf{k}$ that fall within a shell of radius $|\mathbf{k}| = k$. This results in the 1D power spectrum:

\begin{equation}
    P(k) = \sum_{|\mathbf{k}| = k} \left( |\tilde{w}_x(\mathbf{k})|^2+ |\tilde{w}_y(\mathbf{k})|^2+|\tilde{w}_z(\mathbf{k})|^2 \right).
\end{equation}

To avoid boundary effects due to the accumulation of magnetic flux on the edges of the box, we applied to both velocity and vorticity fields a window function along the $x-$direction. We adopted the Kaiser-Bessel window \citep{HarrisKaiserBessel1978}, defined as

\begin{equation}
    W(n) = I_0 \left( \alpha \sqrt{1 - \frac{4n^2}{(M-1)^2}} \right) \bigg / I_0(\alpha),
\end{equation}

\noindent
with

\begin{equation}
    - \frac{M-1}{2} \le n \le \frac{M-1}{2},
\end{equation}

\noindent
where $I_0$ is the modified zeroth-order Bessel function, $\alpha$ is a shape parameter, and $M=1025$ is the number of points in the output window. We have adopted $\alpha = 6.0$, corresponding to a function similar to a Hann window. Once we adopted periodic boundary conditions at the $z-$boundaries, no windowing is necessary along this direction.

Figure \ref{fig:powerspec_vorticity} shows the power spectrum for the velocity ($\mathbf{v}$, left) and vorticity ($\omega = \nabla \times \mathbf{v}$, right) fields at five time-steps depicted in Figures \ref{fig:2Dcuts_currdens_S1e5_bt2_P0.5_k128} and \ref{fig:2Dcuts_vorticity_S1e5_bt2_P0.5_k128}. The dashed black lines represent the Kolmogorov power spectrum $\mathcal{P} (k) \propto k^{-8/3}$ (velocity) and $\mathcal{P} (k) \propto k^{-2/3}$ (vorticity).




We observe that the power spectra for both velocity and vorticity (Fig. \ref{fig:powerspec_vorticity}) align with the Kolmogorov spectral slopes (dashed lines), indicating that the system transitions into a turbulent state following the initial external perturbation. 
This turbulence evolves freely after the initial energy input, without continuous external driving. As shown in Fig. \ref{fig:powerspec_vorticity} (right), the spectrum of the decaying turbulence develops  steeper slopes 
particularly at small scales (high $k$), where energy dissipates  more rapidly. While the spectrum in the inertial range maintains a $k^{-2/3}$ scaling, the progressive decay leads to increasingly steeper slopes at smaller scales.


Although the curves exhibit similarities for $t \ge 0.5 \, t_A$, the increase in the total power observed across all scales throughout  the temporal evolution (from $t=0.1 \, t_A$ up to $2.5 \, t_A$) provides clear evidence of turbulence generation. In the final stage of the simulation ($t=3.0 \, t_A$), we observe  saturation in the total energy density of the turbulence (compare the red and purple lines in Fig. \ref{fig:powerspec_vorticity}), which also influences the reconnection rate. This aspect will be discussed in the following sections.

\subsection{Dependence on the plasma$-\beta$}

By studying the statistics of reconnection-driven turbulence, \cite{kowal2017statistics} noticed a weak dependence of the reconnection rate with the plasma parameter $\beta$, the ratio between the thermal and magnetic pressures:

\begin{equation}
    \beta = \frac{p_{\text{th}}}{p_{\text{mag}}}= \frac{2 \rho c_s^2}{B^2}, 
\end{equation}

\noindent
where $c_s$ is the isothermal speed of sound and $B$ is the amplitude of the magnetic field, which in code units is normalized by $1/\sqrt{4 \pi}$.

Indeed, \cite{kowal2017statistics} demonstrated a decrease in the reconnection rate with $\beta$, while maintaining constant the magnetic pressure ($B^2$ constant). Consequently, this result indicates a  correlation between $V_{\text{rec}}$ and the plasma's sound speed or compressibility, aspects overlooked in the original theory of turbulent reconnection by \cite{lazarian1999reconnection}.

\begin{figure}[t]
    \centering
    \includegraphics[width = 0.47 \textwidth]{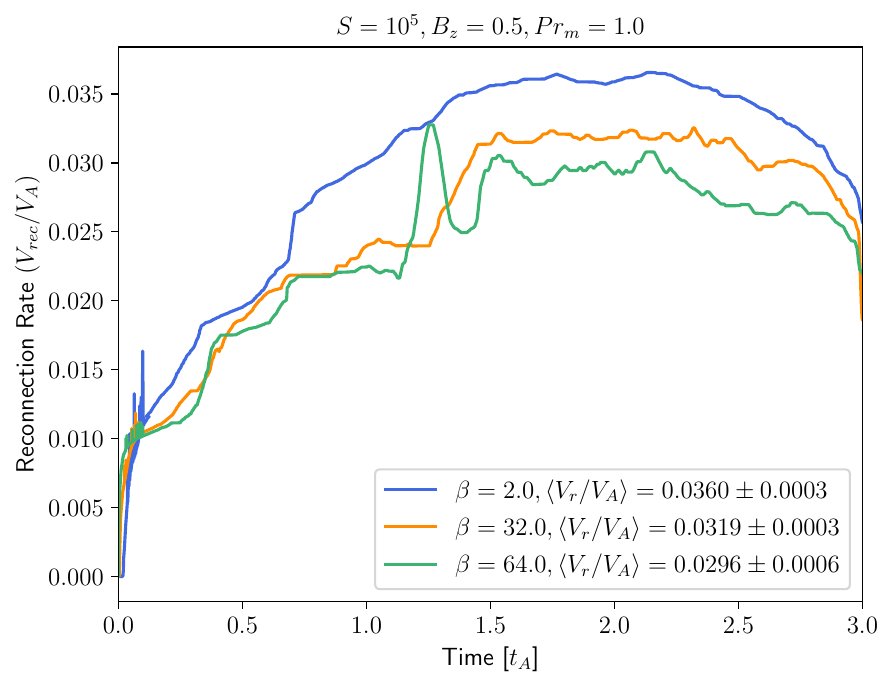}
    \caption{Time evolution of the reconnection rate for the simulations with $S = 10^5$, $\text{Pr}_m = 1.0$, $P_\text{inj} = 0.1$, $k_\text{inj}=128$, and $\beta = \{2.0, 32.0, 64.0\}$. }
    \label{fig:vrec_betas_time_strong_filter}
\end{figure}

In Figure \ref{fig:vrec_betas_time_strong_filter} we show the time evolution of the reconnection rate for the simulations with $S = 10^5$, $\text{Pr}_m=1$, $P_\text{inj} = 0.1$, $k_\text{inj}=128$ and  three different values of $\beta$ ($2.0, 32.0, 64.0$). 
{To enhance clarity between the curves, we have applied a strong filter to smooth them, using a median filter with a width of around 3000 points for computing the time derivatives of magnetic flux. This  corresponds to approximately $0.03 \, t_A$ ($\Delta t_{\text{step}} \sim 10^{-5}$) in data filtering.} From Fig. \ref{fig:vrec_betas_time_strong_filter} we notice a dependence of $V_\text{rec}/V_A$ on $\beta$, with the reconnection rate decreasing as $\beta$ increases.

We also notice from Figure \ref{fig:vrec_betas_time_strong_filter} that,
 for times $t \gtrsim 2.5 \, t_A$, the reconnection rate starts to decrease, which may be caused by the exhaustion of the magnetic flux or by boundaries effects - with the accumulation of magnetic flux along the $x-$boundaries. Therefore, the average rates shown in this figure are calculated between $t=1.5 \, t_A$ and $2.5 \, t_A$, where the reconnection rate reaches a quasi-steady state, and we find $\langle V_\text{rec}/V_A \rangle = 0.0360(3), 0.0319(3),  0.0296(6)$ for $\beta = 2, 32, 64$, respectively.

\begin{figure}[t]
    \centering
    \includegraphics[width = 0.47 \textwidth]{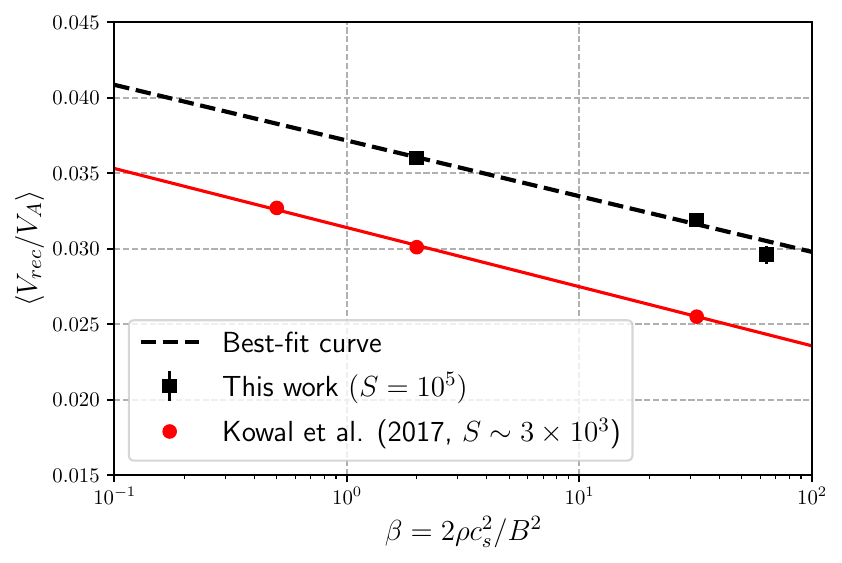}
    \caption{Time  averaged reconnection rate for different values of plasma$-\beta$. The black dashed line is the best-fit curve, $\langle V_{\text{rec}}/V_A \rangle = 0.0372 - 0.0016 \log \beta$. 
    The red dots and line represent the results from self-generated turbulent reconnection performed by \cite{kowal2017statistics}. 
     }
    \label{fig:vrec_betas_1sigma}
\end{figure}

 Figure \ref{fig:vrec_betas_1sigma} shows the time averaged reconnection rate $\langle V_{\text{rec}}/V_A \rangle $, between $1.5 \, t_A\le t \le 2.5 \, t_A$, as a function of $\beta$. The dashed black line represents the best-fit curve of $\langle V_{\text{rec}}/V_A \rangle = a + b \log \beta$, with $a = 0.0372(4)$ and $b=-0.0016(1)$. The red dots represent the results obtained by \cite{kowal2017statistics}, where turbulence is self-generated in the reconnection layer with Lundquist number of $S \approx 3 \times 10^3$, while the red solid line is the fit of their data, given by $\langle V_\text{rec} / V_A \rangle = 0.0314 - 0.0017 \log \beta$.

The reconnection rates derived in this study consistently exhibit slightly higher values (approximately $0.006 \, V_A$ higher) compared to those reported by \cite{kowal2017statistics}. This discrepancy is likely due to differences in the initial perturbations implemented in the simulations and the onset conditions of magnetic reconnection. 
In our setup, small-scale multi-mode perturbations are injected from $t=0$ to $0.1 \, t_A$, whereas \cite{kowal2017statistics} introduce an initial velocity perturbation with randomly distributed directions. 
Additionally, the overall onset of reconnection differs between the two studies. In our case, reconnection occurs through the merging of two flux tubes with closed boundaries along the $x$ and $y$ directions. In contrast, \cite{kowal2017statistics} use an initial magnetic field configuration that is antiparallel along the $x$ direction, with a discontinuity at $y = 0$ in a larger simulation box, and open boundary conditions along the $y$ direction.
Despite these differences, both studies observe a similar weak correlation between the reconnection rate and the plasma parameter $\beta$. 
By lowering the plasma$-\beta$ while keeping the magnetic field constant, the sound speed ($c_s$) decreases, leading to a higher Mach number ($M = v_\text{turb}/c_s$). Consequently, turbulence becomes more dominant, intensifying the wandering of magnetic field lines and enhancing the turbulent reconnection rate.


It is  important to remark that
in the simulations depicted above, the Lundquist number was set to be $S=10^5$, therefore, much higher than the one employed in the numerical simulations of \cite{kowal2017statistics}, of the order of $S \approx 3 \times 10^3$. Still, the reconnection rates calculated from the self-generated turbulence regime by \cite{kowal2017statistics} are of the same order as the ones we obtained, of $V_{\text{rec}}/V_A \sim 0.025 - 0.035$, reinforcing the prediction of \cite{lazarian1999reconnection} theory of reconnection, where the turbulence makes the reconnection rate independent on the Ohmic resistivity and the Lundquist number.

\subsection{Dependence on the magnetic Prandtl number}

The dependence of the reconnection rate on the viscosity was theoretically studied and numerically tested by several authors \citep{park1984reconnection, kowal2012visc, Comisso_2015, Jafari_2018}. Considering the energy balance and mass conservation along the current sheet and the topology of the magnetic field lines, \cite{park1984reconnection} found a dependence of the reconnection rate on the magnetic Prandtl number ($\text{Pr}_m = \nu / \eta$) for the Sweet-Parker and the Petschek models, given by $V_\text{rec, SP} \propto (1 + \text{Pr}_m)^{-1/4}$ and $V_\text{rec, Pet} \propto (1 + \text{Pr}_m)^{-1/2}$, respectively. 

By solving the Taylor problem in the plasmoid-unstable regime, \cite{Comisso_2015} obtained the scaling of $V_\text{rec, plm} \propto (1 + \text{Pr}_m)^{-1/2}$ for this regime, which is  compatible with their 2D numerical simulations of current sheets for a range of  high magnetic Prandtl numbers $(20 \lesssim \text{Pr}_m \lesssim 50)$.

In 3D turbulent flows obtained from numerical studies of the Lazarian-Vishniac theory for turbulence driven models with $S \sim 10^3$, \cite{kowal2012visc} found a weaker dependence of $V_\mathrm{rec}$ with the Prandtl number, $V_\mathrm{rec} \sim \mathrm{Pr}_m^{-1/4}$. Later, \cite{Jafari_2018} demonstrated that, for high viscosity  ($\nu \gg \langle y^2\rangle$, where $\langle y^2\rangle$ is the mean quadratic separation of the magnetic field lines), the reconnection rate should scale with the Prandtl number as $V_\text{rec} \propto \exp (\text{Pr}_m^{-1/2})$. On the other hand, these authors argue that the condition for viscosity to be important, given by $\langle y^2\rangle^{1/2} \lesssim \ell_{d\nu}$ (where $\ell_{d \nu}$ is the Kolmogorov viscous damping scale), can be written in terms of the Reynolds number ($\text{Re}$) as

\begin{equation}
    \text{Re}^{1/2} < \left( \dfrac{\ell_\parallel}{L_x} \right) \left[ 1 + \ln (\text{Pr}_m)\right].
\end{equation}

\noindent
where $\ell_\parallel$ is the parallel scale of the turbulent eddies with respect to
the mean magnetic ﬁeld. Therefore, the width of the turbulent current sheet is unaffected by the viscosity unless the magnetic Prandtl number is exponentially larger than the Reynolds number, and this condition is
unlikely to be satisﬁed both in astrophysical environments and in numerical simulations. In other words, if the ﬁeld line separation is totally in the inertial range ($\langle y^2\rangle^{1/2} > \ell_{d\nu}$) viscosity has a negligible effect on the turbulent reconnection rate.

\begin{figure}[t]
    \centering
    \includegraphics[width = 0.48 \textwidth]{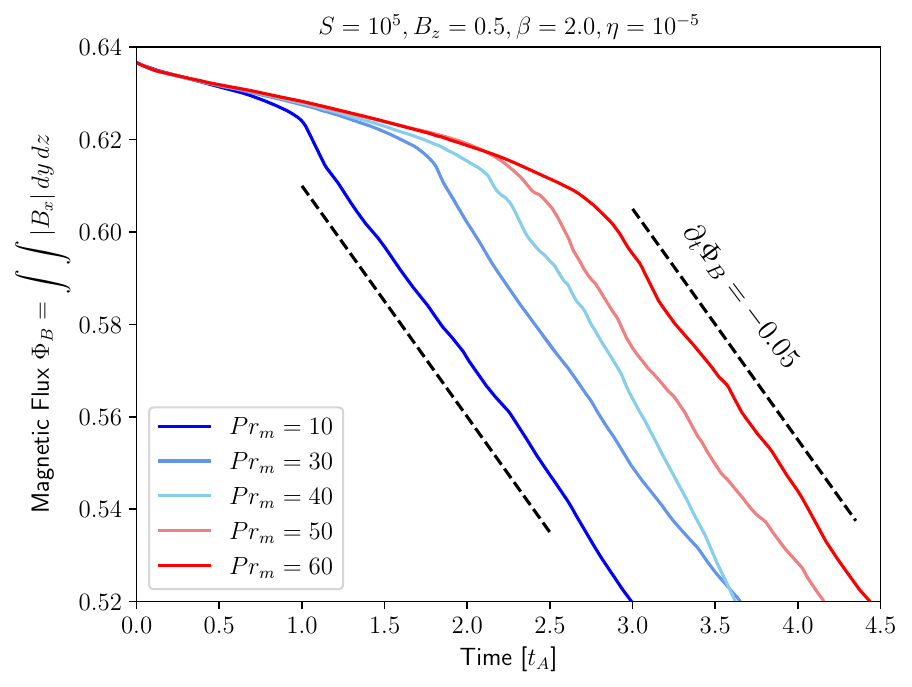}
    \caption{Time evolution of the magnetic flux $\Phi_B$, calculated across the plane $x=0$, for different magnetic Prandtl numbers. The two dashed black lines correspond to a constant reconnection rate of $V_\text{rec}/V_A = -\partial_t \Phi_B = 0.05$.}
    \label{fig:Bflux_Prandtl}
\end{figure}

The simulated models depicted in Figures \ref{fig:3D_visu_t0_t2} to \ref{fig:vrec_betas_1sigma}, had magnetic Prandtl number $\text{Pr}_m=1$.  Aiming to test these analytical findings we also performed 3D MHD numerical simulations of current sheets for high magnetic Prandtl numbers, ranging from $\text{Pr}_m = 10$ to $60$. In Figure \ref{fig:Bflux_Prandtl} we show the time evolution of the magnetic flux $\Phi_B$ (calculated across the plane $x=0$) for the range of $\text{Pr}_m$ mentioned. All the simulations were performed with resistivity $\eta = 10^{-5}$, or equivalently $S=10^5$, $\beta = 2.0$, and guide field $B_z = 0.5$. The initial perturbation was set in the system by injecting forced turbulence with power $P_\text{inj} = 0.1$ and $k_\text{inj}=128$, from $t=0$ up to $0.1 \, t_A$. 

Based on Fig. \ref{fig:Bflux_Prandtl}, we can observe that the higher the viscosity, the longer it takes for the turbulence to become self-sustained and enhance the magnetic flux decay.
In simulations with $\text{Pr}_m = 1$ (Figs. \ref{fig:powerspec_vorticity} and \ref{fig:vrec_betas_time_strong_filter}), we find that the initial perturbation rapidly triggers the instabilities, which in turn drives turbulence almost immediately. 
Conversely, for cases with larger $\text{Pr}_m$, the higher viscosity slows down turbulence development. Increased viscosity reduces the inertial range, suppressing small-scale structures and causing both the instabilities and turbulence to grow more gradually (slowly).

\begin{figure}[t]
    \centering
    \includegraphics[width = 0.48 \textwidth]{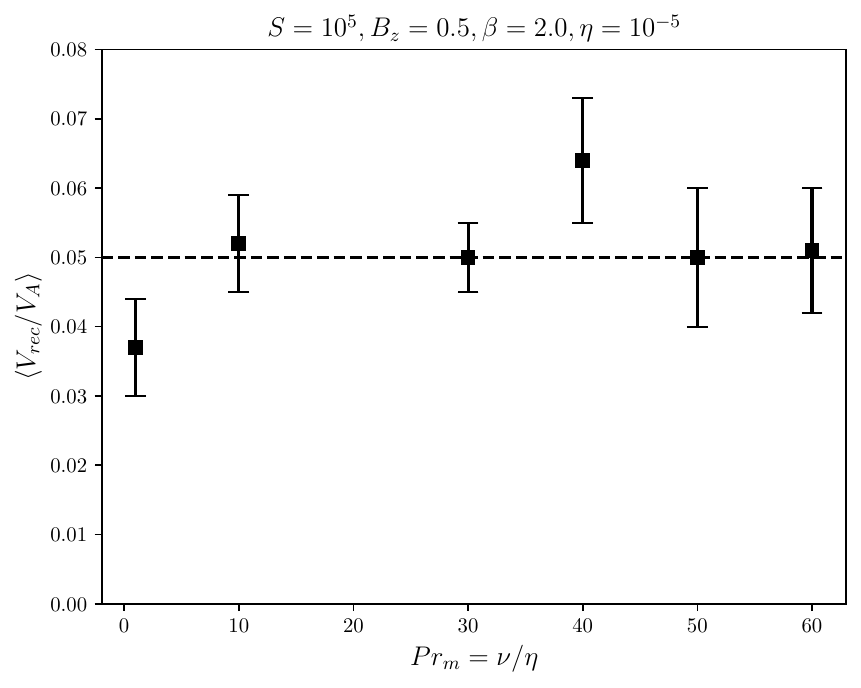}
    \caption{Time averaged reconnection rates for different Prandtl numbers for the simulations with $\eta = 10^{-5}$, $\beta = 2.0$ and $B_z = 0.5$.}
    \label{fig:Average_Vrec_Prandtl}
\end{figure}

In Fig. \ref{fig:Bflux_Prandtl}, once the turbulence is fully developed in the 3D domain, all the models converge to a constant reconnection rate of $V_\text{rec}/V_A = -\partial_t \Phi_B \sim 0.05$ (represented by the two dashed black lines). By averaging in time the reconnection rates for these five models, we  obtain Fig. \ref{fig:Average_Vrec_Prandtl}, which also includes the simulated model  with $\text{Pr}_m = 1$ (Fig. \ref{fig:vrec_betas_time_strong_filter}).

We notice from Figs. \ref{fig:Bflux_Prandtl} and \ref{fig:Average_Vrec_Prandtl} that, for the case of $\text{Pr}_m = 40$, the magnetic flux decays faster, increasing the average reconnection rate to $\langle V_\text{rec}/V_A \rangle = 0.064(9)$. On the other hand, we should emphasize that all the points in Fig. \ref{fig:Average_Vrec_Prandtl} match to the constant value of $\langle V_\text{rec}/V_A \rangle \sim 0.05$ within a confidence interval of $2\sigma$. More significant is the fact that there is no evidence of an explicit dependence of the reconnection rate on the magnetic Prandtl number, or equivalently on the viscosity, for the turbulent magnetic reconnection, as stated by \cite{Jafari_2018}, and in agreement with the theory proposed by \cite{lazarian1999reconnection}.

\begin{figure}[t]
    \centering
    \includegraphics[width = 0.48 \textwidth]{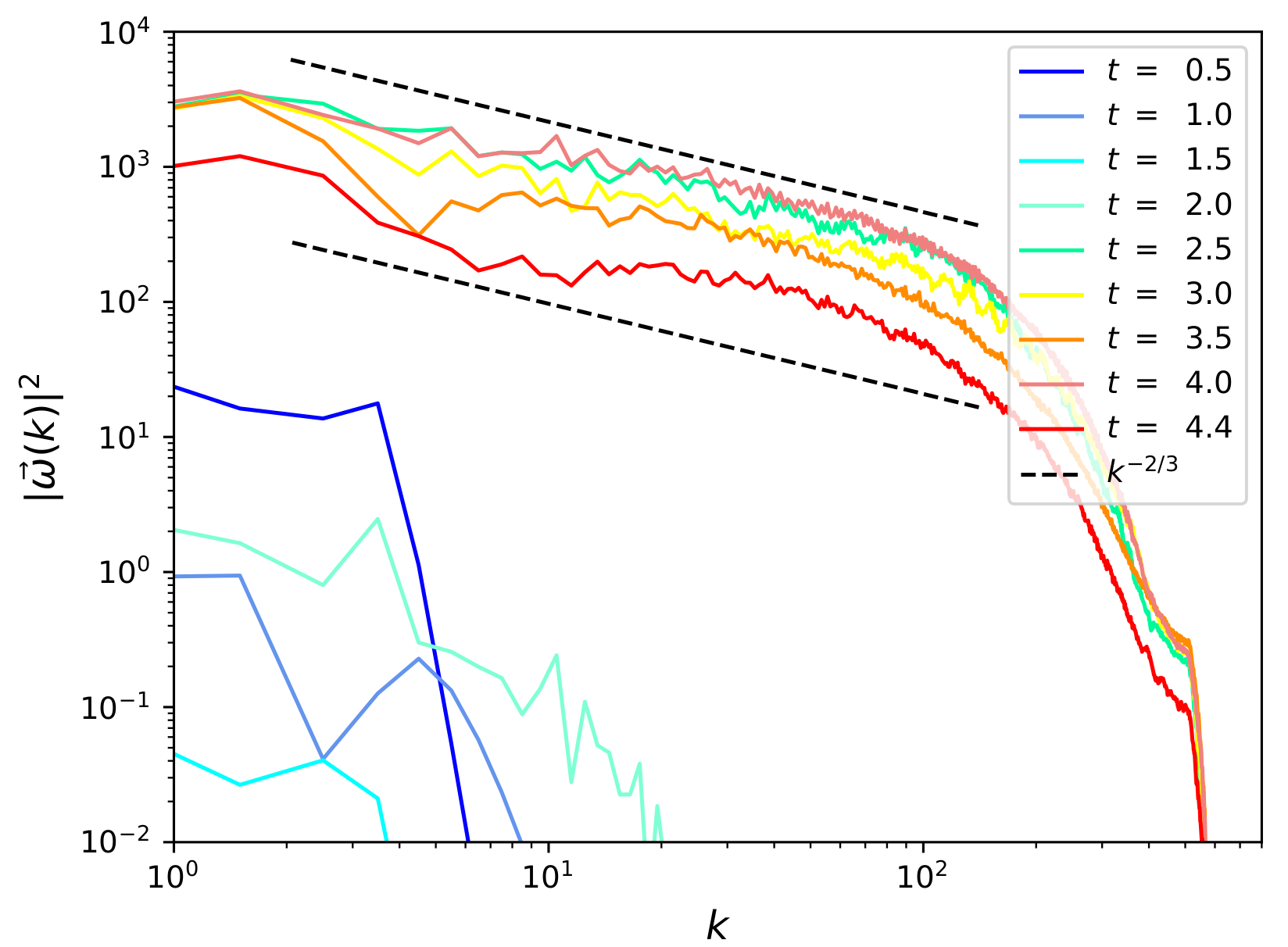}
    \caption{Power Spectra of the vorticity field in the $xz-$plane (current sheet mid-plane) at different timesteps of the simulation with $S = 10^5$, $\beta=2.0$ and $\text{Pr}_m = 50$. The black dashed lines represents the 2D Kolmogorov turbulent power spectrum of $\mathcal{P}(k) \propto k^{-2/3}$. }
    \label{fig:power_spec_vort_Prm50}
\end{figure}

\begin{figure*}[t]
    \centering
    \includegraphics[width = 0.98 \textwidth]{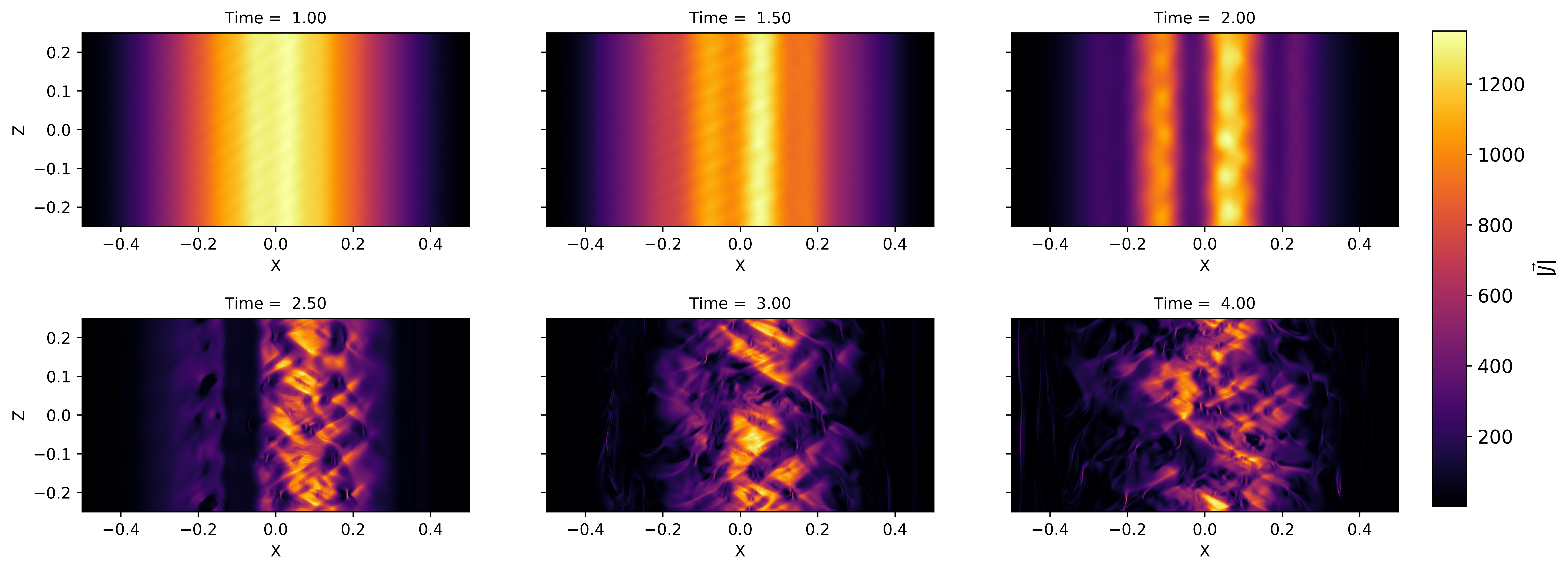}
    \caption{Colormaps of 2D cuts ($xz-$plane) of the current density magnitude $(|\mathbf{J}| =  |\nabla \times \mathbf{B}|)$ at different times for the simulation with $S = 10^5$, $\beta = 2.0$, and $\text{Pr}_m = 50$, 
    and initial small-scale perturbation with $k_\text{inj} = 128$ and $P_\text{inj} = 0.1$.
    }
    \label{fig:combined_2Dcuts_jj_xz_Pr50}
\end{figure*}

In Figure \ref{fig:power_spec_vort_Prm50} we show the power spectrum evolution  of the vorticity ($|\omega (k)|^2$) for the simulation with $\text{Pr}_m = 50$ (light coral line in Fig. \ref{fig:Bflux_Prandtl}). We  notice that for early times, of $t \lesssim 2.0 \, t_A$, where the reconnection rate is slower, the spectrum does not follow the turbulent Kolmogorov scaling of $\mathcal{P}(k) \propto k^{-2/3}$. On the other hand, for times $t \gtrsim 2.5 \, t_A$, when the reconnection rate becomes fast and nearly constant around $\langle V_\text{rec}/V_A \rangle \sim 0.05$, the vorticity spectra  calculated along the current sheet ($xz-$plane) become consistent with the turbulent Kolmogorov scaling, reinforcing the fact that turbulence  is the mechanism responsible for enhancing the reconnection rate. 
The same behavior is observed in all the simulations with different magnetic Prandtl numbers, evidencing the formation of a self-sustained turbulent current sheet as the reconnection rate becomes fast.

\subsection{Turbulence versus tearing mode instability}


In this section, we contrast the results from our simulations with high Lundquist numbers $(S = 10^5)$ (where the tearing mode (plasmoid)  instability may occur) under conditions with and without initial external perturbation.

In the previous Section, we have found that the reconnection rate is enhanced as the turbulence is developed in the 3D domain. 
The initial perturbation in the system, coupled with the high Lundquist number of $S = 10^5$, is expected to render the system unstable to the tearing mode instability. This instability is responsible for a  reconnection rate of approximately $\sim 0.01 \, V_A$, a universal rate  associated with the formation of a chain of plasmoids (magnetic islands) in 2D current sheets in MHD simulations \citep[see e.g.,][]{loureiro2007instability, bhattacharjee2009fast, huang2010scaling}.

Figure \ref{fig:combined_2Dcuts_jj_xz_Pr50} shows 2D cuts of the current density magnitude along the $xz-$plane at different time steps of the simulation for $\text{Pr}_m = 50$. We can notice the development of the tearing instability and the formation of flux ropes in the current sheet during the early times of the simulation, between $t = 1.0 - 2.0 \, t_A$, which represents the 3D view of the reconnected layers. Plasmoids should represent 2D cuts of such flux ropes in the $xy-$plane.

\begin{figure*}[t]
    \centering
    \includegraphics[width=0.7\textwidth]{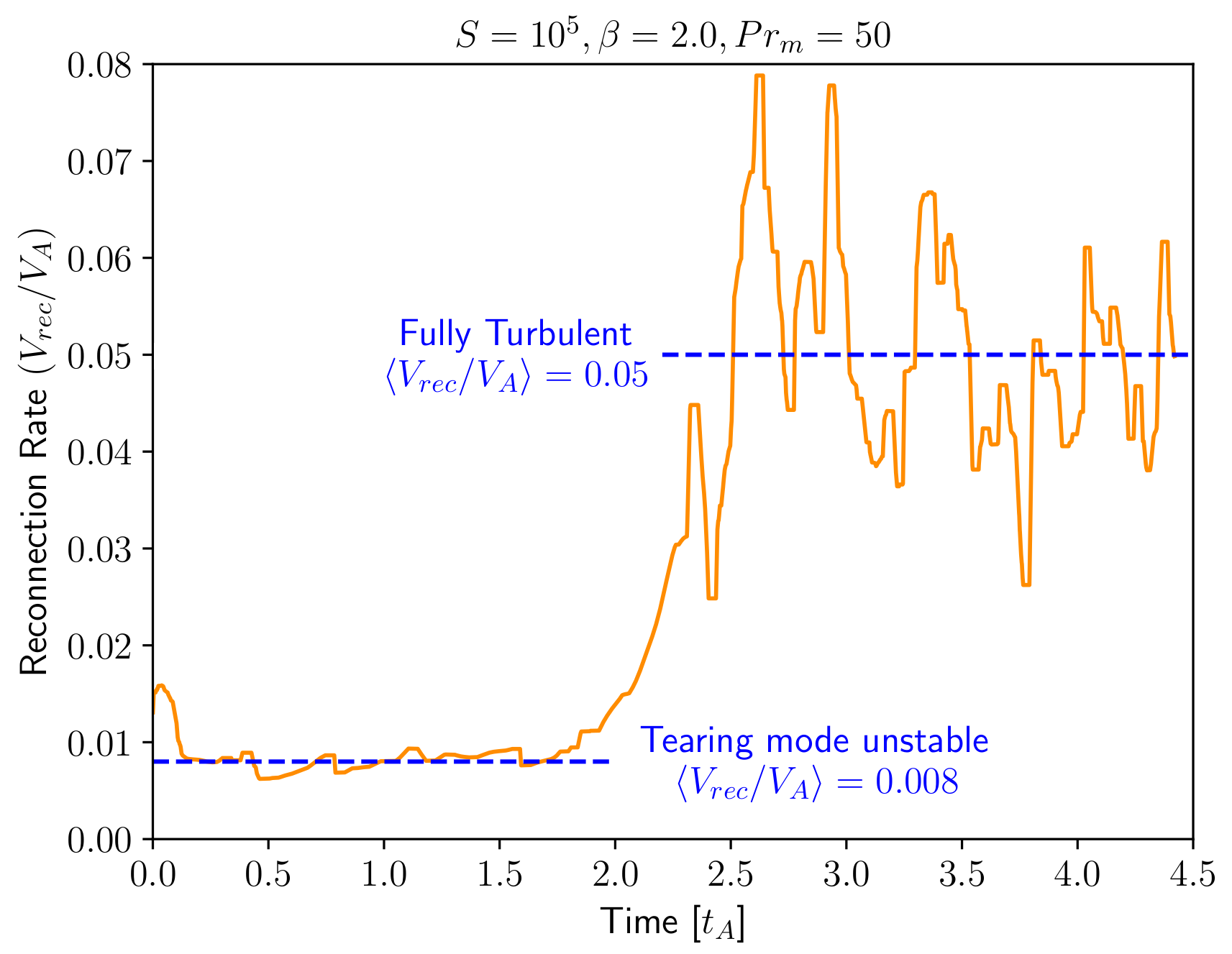}
    \caption{Time evolution of the reconnection rate for the simulation with $S = 10^5$, $\beta = 2.0$, $\eta = 10^{-5}$ and $\text{Pr}_m=50$. The blue dashed lines represent the average reconnection rates of $\langle V_\text{rec}/V_A \rangle = 0.008$ (tearing mode unstable regime)
    and $0.05$ (fully turbulent regime). 
    }
    \label{fig:vrec_plasmoid_turb}
\end{figure*}

By calculating the time derivative of the magnetic flux across the plane perpendicular to the current sheet (Eq. \ref{eq:vrec_kowal09_thiswork}), we show the reconnection rate in Figure \ref{fig:vrec_plasmoid_turb},  for the same simulation of Figure \ref{fig:combined_2Dcuts_jj_xz_Pr50} ($\text{Pr}_m = 50$). The initial average rate of $\langle V_\text{rec}/V_A  \rangle = 0.008(1)$ is attributed solely to the tearing instability since in this stage we do not observe the development of turbulence  in the power spectrum yet (see Figure \ref{fig:power_spec_vort_Prm50}). This value is compatible with the extensively tested and so-called standard
rate of $0.01 \, V_A$ in this regime for MHD models. But, once turbulence is fully developed in the current sheet, for times $t \gtrsim 2.5 \, t_A$ (bottom panels of Fig. \ref{fig:combined_2Dcuts_jj_xz_Pr50}), the reconnection rate is enhanced by a factor of 5, with an average rate of $\langle V_\text{rec}/V_A  \rangle = 0.050(12)$, reaching peaks of $V_\text{rec} \sim 0.08 \, V_A$. This fast reconnection rate is compatible, e.g., with the ones obtained from local simulations of self-generated turbulence performed by \cite{kowal2017statistics} and from global simulations of accretion flows \citep{Kadowaki_2018} and relativistic jets \citep{singh2016spatial, Kadowaki_2021, Medina-Torrejón_2021, Medina-Torrejón_2023}, where turbulence is driven by the magnetorotational instability (MRI) and the current-driven kink instability (CDKI), respectively. 

It is interesting to note that we obtain the same reconnection rate evolution as shown in Figure \ref{fig:vrec_plasmoid_turb} when using the method described by \cite{Huang_2016} to evaluate the reconnection rate (see Appendix \ref{appendix:rec_rate}  and Figure  \ref{fig:compar_vrec_BH16}).

Additionally, we reproduced  Figure \ref{fig:vrec_plasmoid_turb} using the initial configuration implemented by \cite{Huang_2016}, which features an initial constant density  and a non-uniform guide field $B_z$, and  have obtained a very similar result   (see  Figure \ref{fig:dif_initial_cond_BZ_dens_const} in Appendix \ref{appendix:dif_initial_cond}).

We can also compare the models with and without initial forcing 
by applying other different methods to calculate the reconnection rate. Using mass conservation, the total mass flux that enters the reconnection layer along length $L$ must equalize the flux that leaves the current sheet of thickness $\delta$ with the Alfvén speed. Therefore, we can approximate the inflow velocity as $v_{in} L \approx V_A \delta$, yielding the reconnection rate

\begin{equation}
    \frac{V_{\text{rec}}}{V_A} \approx \frac{v_{in}}{V_A} \approx \frac{\delta}{L}. \label{eq:vrec_delta_L}
\end{equation}


\begin{figure}[h!]
    \centering
    \includegraphics[width = 0.45 \textwidth]{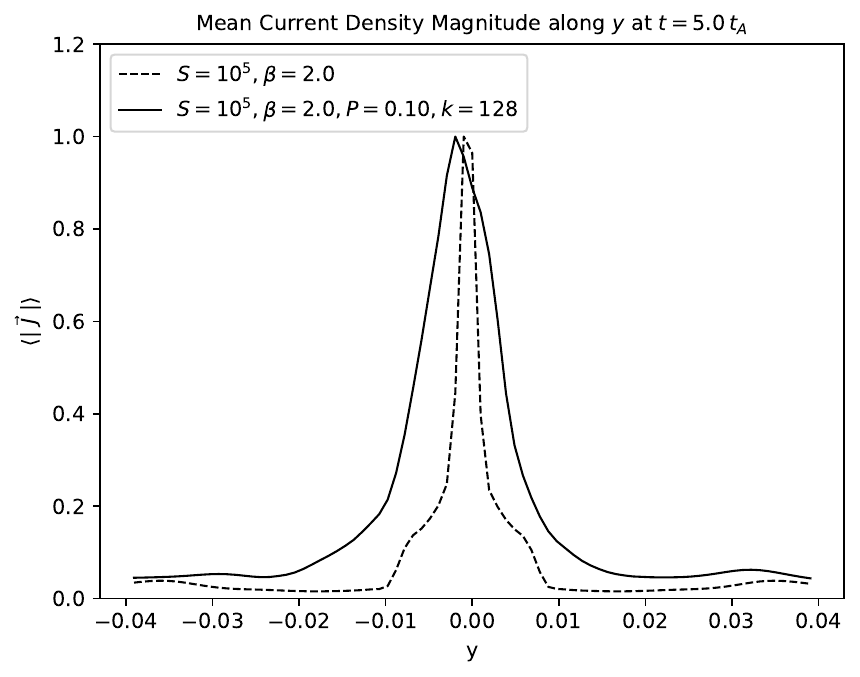}
    \caption{Comparison of the mean current density magnitude profile along the $y-$axis between the model with initial external perturbation where turbulent develops (solid) and the one without it, where tearing instability develops (dashed line). Both simulations have $S = 10^5$, $\beta = 2.0$, and $\text{Pr}_m = 1$.
    }
    \label{fig:profile_meanJ_along_y}
\end{figure}

\begin{figure*}[!htb]
    \centering
    \includegraphics[width = 0.48 \textwidth]{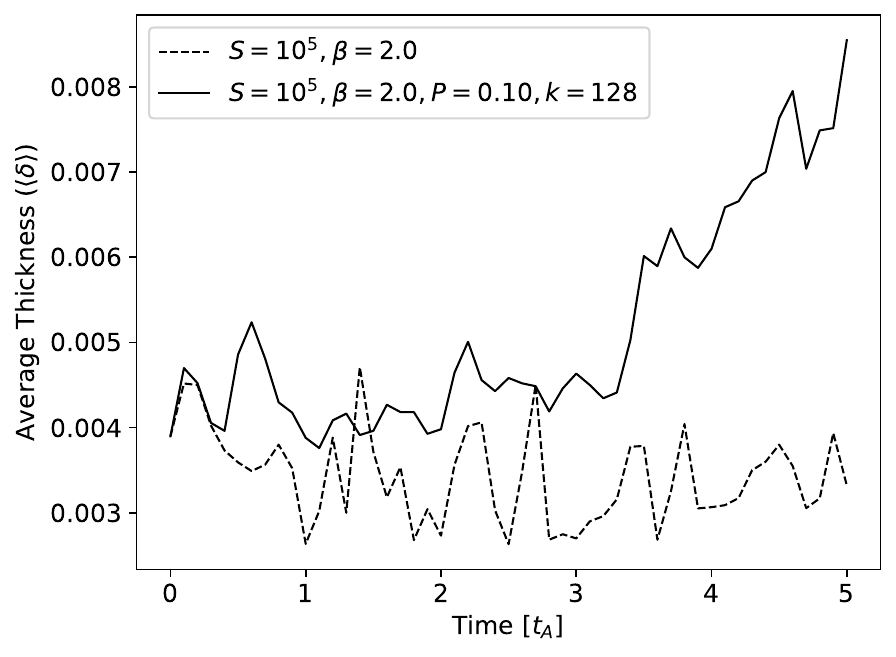}
    \includegraphics[width = 0.48 \textwidth]{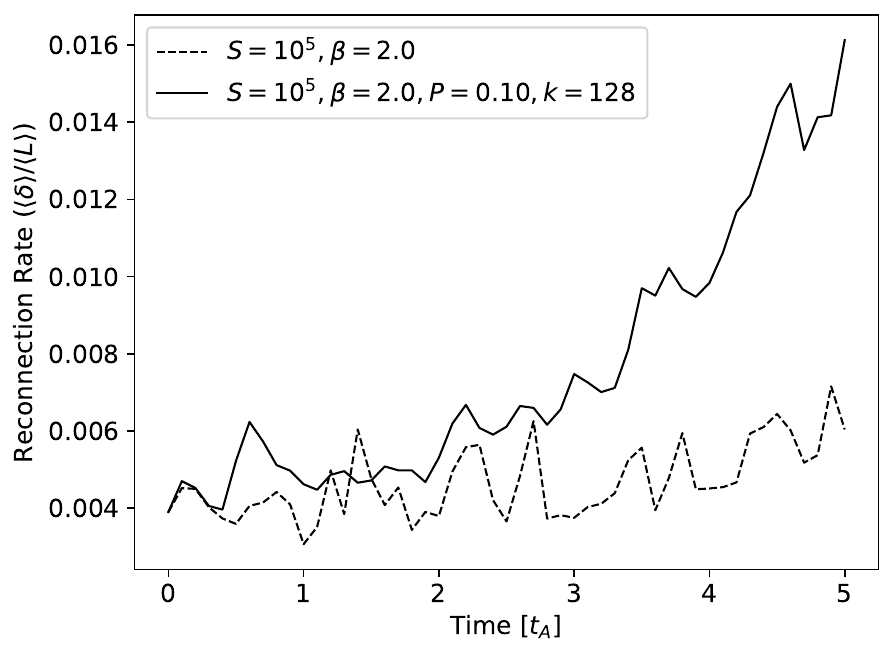}
    \caption{\textbf{Left:} Average thickness $\langle \delta \rangle$ of the current layer extracted from the FWHM of the current density magnitude profile for the turbulent (solid) and 
    tearing instability (dashed) cases. \textbf{Right:} An estimate of the reconnection rate for these two simulations, as measured from the ratio $\langle \delta \rangle / \langle L \rangle$. The models  are the same as those in Figure \ref{fig:profile_meanJ_along_y}.    }
    \label{fig:avg_delta_vrec}
\end{figure*}

\begin{figure*}
    \centering
    \includegraphics[width = 0.999 \textwidth]{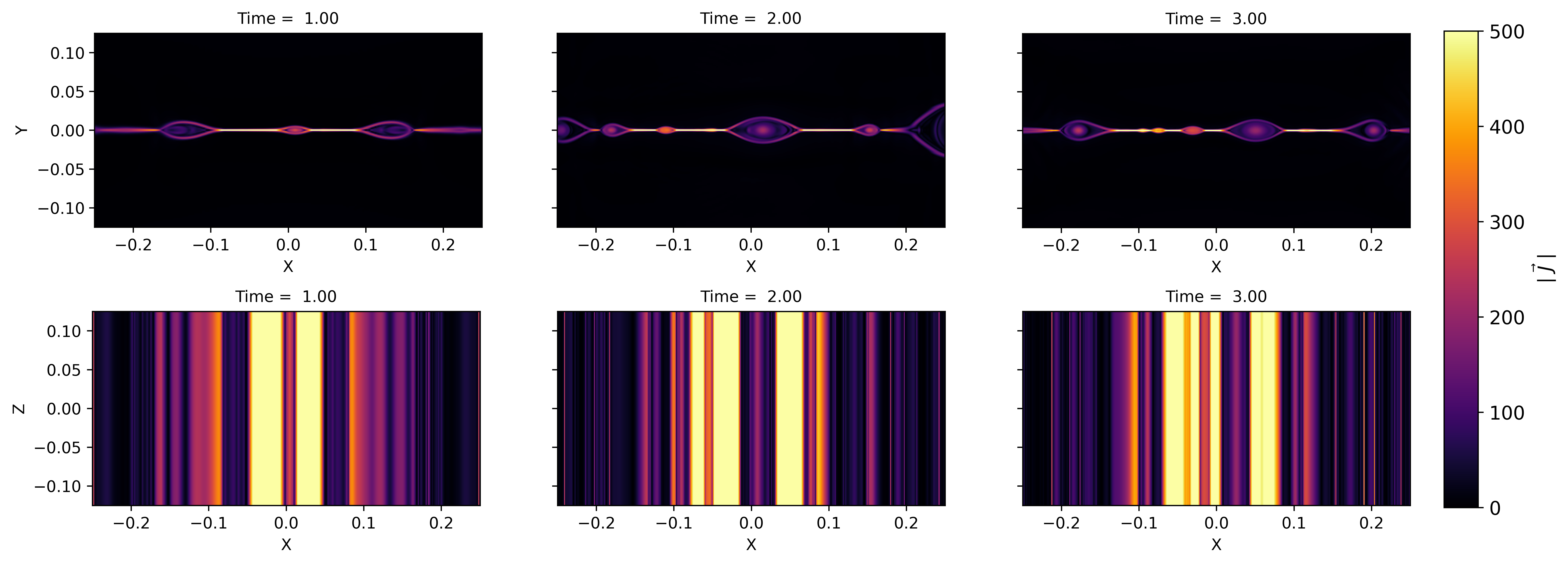}
    \caption{Colormaps of 2D cuts along the planes $z=0$ (top) and $y=0$ (bottom) of the current density magnitude at different times for the simulation with $S = 10^5$, $\beta = 2.0$, and $\text{Pr}_m = 1$, without any initial perturbation.}
    \label{fig:combined_jj_xy_xz_noturb}
\end{figure*}

\begin{figure*}
    \centering
    \includegraphics[width = 0.45 \textwidth]{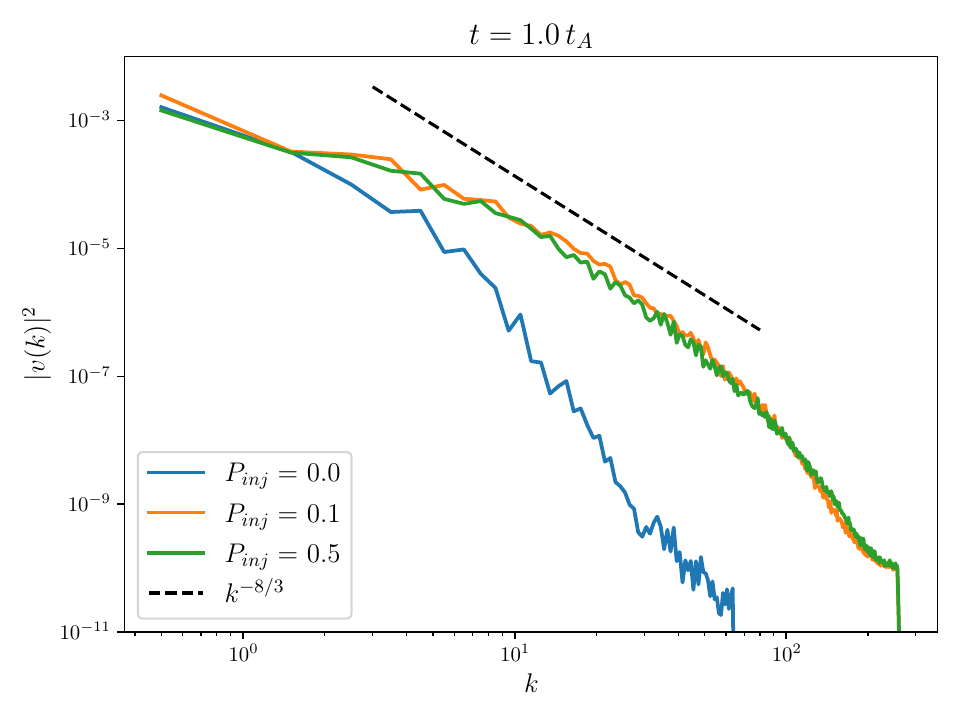}  \includegraphics[width = 0.45 \textwidth]{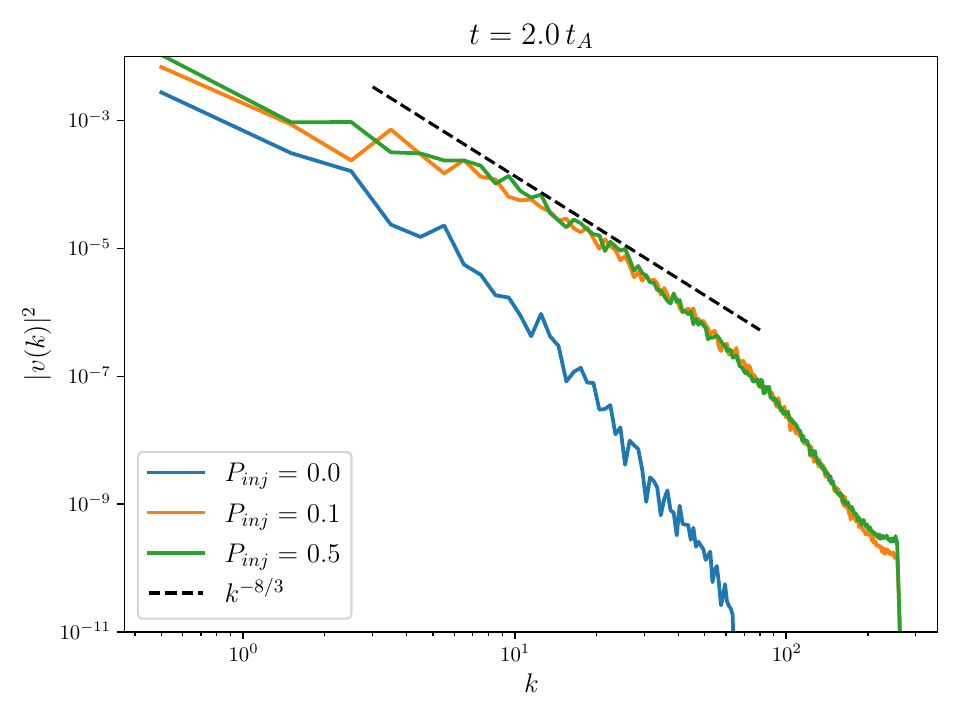}
    \caption{Power spectra of the velocity field at $t = 1.0 \, t_A$ (left) and $t = 2.0 \, t_A$ (right) for the simulations with $S=10^5$, $\beta =2$, $\text{Pr}_m = 1$, and different injection powers of the initial small-scale perturbation.}
    \label{fig:psepc_pinjs}
\end{figure*}

To estimate the thickness $\delta$ of the current sheet, we plot in Figure \ref{fig:profile_meanJ_along_y} the profile of the normalized mean current density magnitude $\langle |\mathbf{J}| \rangle$ along the $y-$axis for two different simulations. Both simulations have $S = 10^5$, $\beta = 2.0$, $\eta = 10^{-5}$, and $\text{Pr}_m = 1$. The difference lies in the fact that for the simulation represented by the black dashed line, we do not inject any type of perturbation. Conversely, for the simulation represented by the solid line, initial perturbation is injected with $P_\text{inj} = 0.1$ and $k_\text{inj}=128$ from $t=0$ up to $0.1 \, t_A$. At each position $y$, we compute the spatial average of the current density magnitude $\langle |\mathbf{J}| \rangle$ over the region $(x, z) \in [-0.1, 0.1] \times [-0.5, 0.5]$.
We determine the average thickness of the current layer at each time step of the simulation by fitting a Gaussian function to the $\langle |\mathbf{J}| \rangle$ profile and adopt $\langle \delta \rangle = 2 \sqrt{2 \ln 2} \sigma$, the full width at half maximum (FWHM) of the distribution, where $\sigma$ is the standard deviation of the fitted Gaussian. 


We notice from Figure \ref{fig:avg_delta_vrec} (left) that the averaged thickness of the current sheet $\langle\delta  \rangle$ is systematically larger for the turbulent system compared to the case where no external perturbation was driven, and only the tearing mode instability takes place. This is consistent with one of the most significant predictions
of the theory of \cite{lazarian1999reconnection}, which suggests that in the presence of turbulence, the width of the current sheet is a function of turbulence intensity rather than the Ohmic resistivity of plasma.

By accounting the reconnection rate from Eq. (\ref{eq:vrec_delta_L}), as the ratio $\langle\delta  \rangle / \langle L \rangle$,  it is evident that the turbulent regime attains the highest reconnection rate
(Fig. \ref{fig:avg_delta_vrec}, right).
This reaches rates three times higher than the ones generated solely by the tearing mode instability (black dashed line), in the absence of initial external perturbation. We also note that these rates are smaller than the ones obtained from the time derivative of the magnetic flux (compare with Figs. \ref{fig:vrec_betas_time_strong_filter} and \ref{fig:vrec_plasmoid_turb}), which may be related to the approximation of Eq. (\ref{eq:vrec_delta_L}), valid for a 2D laminar flow. But, as a matter of comparison, we still observe a prevalence of turbulence for driving fast reconnection since the inflow velocity estimated from this method for the tearing unstable regime attains values compatible with the resistive reconnection rate of $V_\text{rec}/V_A \approx S^{-1/2} \approx 0.0032$. 

The results of Fig. \ref{fig:combined_jj_xy_xz_noturb} clearly indicate that even in the absence of an initial external perturbation, the tearing mode instability still grows. This is likely triggered by numerical errors, which act at effective scales smaller than our imposed perturbation at \( k = 128 \). However, the plasmoid-like structures we observe in the \( xy \)-plane of this 3D simulation evolve into flux tube-like structures along the third direction (\( z \)-axis) without developing turbulence. In this case, the plasmoids are thicker than the laminar current layer, but the average thickness of the current sheet does not increase significantly, in contrast to the case with initial external perturbation which leads to turbulence (see Figures \ref{fig:profile_meanJ_along_y}, \ref{fig:avg_delta_vrec} and \ref{fig:psepc_pinjs}).

In Fig. \ref{fig:psepc_pinjs}, we present the power spectra of the velocity field at $t = 1.0 \, t_A$ and $t = 2.0 \, t_A$ for simulations with different injection powers of the initial small-scale perturbation:$P_\text{inj} = \{0.0, 0.1, 0.2\}$. As noted, even in the absence of an initial perturbation ($P_\text{inj} = 0$, blue lines in Fig. \ref{fig:psepc_pinjs}), the tearing instability still emerges (Fig. \ref{fig:combined_jj_xy_xz_noturb}). However, the total energy density of the power spectrum in the undriven case is lower and steeper compared to the cases  with $P_\text{inj} = 0.1$ and $0.5$, where turbulence develops, leading to a slope consistent with the Kolmogorov turbulent spectrum of $\mathcal{P}(k) \propto k^{-8/3}$ (black dashed line).


\section{Discussion} \label{sec:discussion}

Numerical studies of magnetic reconnection have been, for decades, mostly restricted to two-dimensional simulations \citep[see e.g.,][and references therein]{matthaeus1985rapid, biskamp1986magnetic, uzdensky2000two, Samtaney_2009, Daughton_2009_transition, Loureiro2009, bhattacharjee2009fast,huang2010scaling, Loureiro2012, Comisso_2015}. 
The initial topology of the current sheet in 2.5D (2D simulations with guide field) and 3D simulations are the same. On the other hand, the system's evolution is different since, in three dimensions, we do not observe the merging and growth of the magnetic islands, forming huge plasmoids as observed in 2D simulations. On the contrary, in 3D simulations, we have the wandering of the magnetic field lines leading to the development of magnetohydrodynamic turbulence, as shown in Figures \ref{fig:2Dcuts_currdens_S1e5_bt2_P0.5_k128} - \ref{fig:powerspec_vorticity}, \ref{fig:power_spec_vort_Prm50} and \ref{fig:combined_2Dcuts_jj_xz_Pr50}. This scenario aligns with \citet{lazarian1999reconnection} reconnection model.

One decade after \cite{lazarian1999reconnection} proposed the theory of turbulent reconnection, \cite{kowal2009numerical} successfully tested their theory by performing 3D MHD simulations of current sheets with Lundquist numbers up to $S = 3 \times 10^3$ and continuously
injecting turbulence in the domain.

\cite{Huang_2016} also studied magnetic reconnection using 3D MHD simulations, but for higher Lundquist number of $S=2 \times 10^5$ (where tearing mode/plasmoid instability may develop) and in the absence of forced turbulence.
They found similar reconnection rates of $V_\text{rec} \sim 0.008 \, V_A$ in both 2D and 3D simulations, which they attributed to the plasmoid instability \citep{bhattacharjee2009fast}, even noticing the development of self-sustained turbulence. In a more detailed analysis of a 3D MHD simulation with the same setup employed by \cite{Huang_2016}, \cite{beg2022evolution} evidenced three stages of the current sheet with high Lundquist number: the laminar 2.5D phase, the transition phase, and the self-generated turbulent regime, the last one with reconnection rates systematically larger than the ones measured by \cite{Huang_2016}, of $V_\text{rec} \sim 0.02 \, V_A$.

By studying the statistics of reconnection-driven turbulence, \cite{kowal2017statistics} noticed a weak dependence of the reconnection rate with the plasma$-\beta$ parameter, of $\langle V_\text{rec} / V_A \rangle = 0.0314 - 0.0017 \log \beta$, 
for $\beta$ values between 0.5 and 32. These simulations have Lundquist numbers of $S \sim 3 \times 10^3$ where, therefore, the tearing mode instability does not develop. In a more recent study, \cite{Kowal_2020} analyzed current sheets with higher Lundquist numbers of $S = 10^5$ and found that the velocity shear (Kelvin-Helmholtz instability) may dominate over the tearing instability on driving turbulence in stochastic reconnection.

Extending upon these previous studies, we here conducted 3D MHD simulations of current sheets covering a wide parameter space, with different Lundquist numbers ($S = LV_A/\eta$), plasma$-\beta$ parameter ($\beta = 2 \rho c_s^2/B^2$), and magnetic Prandtl number ($\text{Pr}_m = \nu/\eta$).

Over all, our study is more closely aligned with the works of \citet{Huang_2016}, 
\citet {kowal2017statistics}, and 
\citet {beg2022evolution} (see also \citet {Beresnyak_2017}), 
as we do not employ continuous turbulence injection like 
\citet{Loureiro2009} or \citet{kowal2009numerical}.
One key difference between our approach and that of \citet{Huang_2016} and \citet {beg2022evolution} lies in the initial perturbation.
Our initial forcing appears to be a stronger perturbation compared to the random velocity
field noise used by those authors.

\begin{figure*}[t]
    \centering
    \includegraphics[width = 0.99 \textwidth]{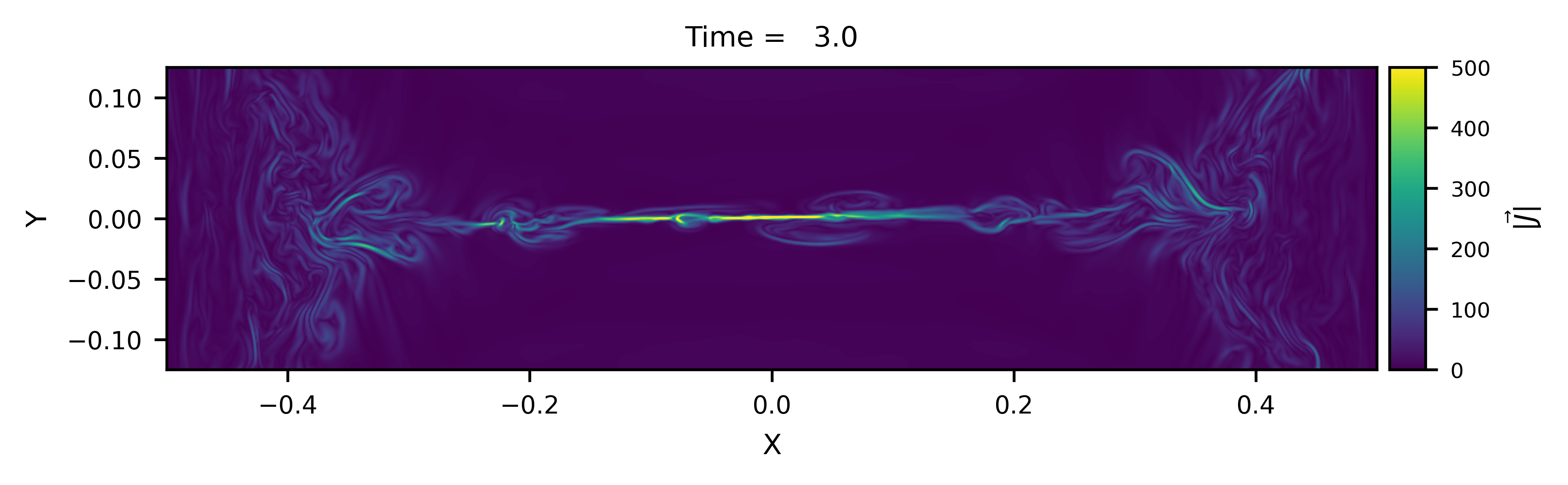}
    \caption{ 2D cut along the plane $z =0$ of the 3D MHD simulation with $S = 10^5, \beta = 2.0, \text{Pr}_m = 1.0$ and initial perturbation of $P_\text{inj} = 0.1$ and $k_\text{inj} = 128$. 
    The time depicted is $t = 3.0 \, t_A$.  In our 3D simulation we do not observe the formation, merging, and growth of the magnetic islands into huge plasmoids which is 
    usually seen in 2D MHD simulations \citep[e.g.,][]{bhattacharjee2009fast, huang2010scaling}.  For a complete view of the entire evolution, please refer to the movie available in the online version of this paper.
    }
    \label{fig:jmag_2D_3D}
\end{figure*}

Covering different Lundquist numbers ranging from $S=10^3$ to $10^6$, in the absence of any initial disturbance, we note the reconnection rate's dependency on $S$, following the Sweet-Parker relation for small Lundquist numbers, as expected for a laminar current sheet. Upon surpassing values of $S>S_c \approx 10^4$, the current sheet becomes prone to tearing mode instability, resulting in the reconnection rate attaining the standard
value of $0.01 V_A$. These rates are compatible with the ones obtained from 2D and 3D MHD simulations of \cite{bhattacharjee2009fast, huang2010scaling, Huang_2016}. 

On the other hand, in our simulations with   injection of small-scale perturbations with $k_\text{inj} =128$ and $P_\text{inj} = 0.1$, 
for a short period of time (up to $t = 0.1 \, t_A$), we observe an increase in the reconnection rate, reaching maximum values nearly an order of magnitude higher compared to those recorded in simulations where fast reconnection is solely driven by tearing mode instability ($\sim 0.008 \, V_A$,  see Figures \ref{fig:Vrec_depS_beta} and \ref{fig:vrec_plasmoid_turb}). 
The initial perturbation triggers 
tearing-like instability, which subsequently drives turbulence, and this leads to the fastest reconnection rate.
We have obtained values consistent with the reconnection rates measured by \cite{kowal2017statistics}, where turbulence 
is self-generated by reconnection, reaching approximately $V_\text{rec} \sim 0.03 , V_A$. 
However, in the simulations of \cite{kowal2017statistics}, the authors estimate the numerical resistivity as $3\times 10^{-4}$, corresponding, as remarked,  to Lundquist numbers $S \sim 3\times 10^3$.
In contrast, our turbulent simulations feature 
$S=10^5$, demonstrating the reconnection rate's independence from $S$, or the Ohmic resistivity, as  proposed by the turbulent reconnection theory of \cite{lazarian1999reconnection}.


Remarkably, previous 3D MHD studies on the stability of current sheets have shown that MHD instabilities can generate turbulence and drive fast reconnection at rates comparable to those found in our models in the turbulent regime. These studies include not only kink instability \citep[e.g.,][]{singh2016spatial, Kadowaki_2021, beg2022evolution}, but also Kelvin-Helmholtz \citep[e.g.,][]{Kowal_2020}, as well as magneto-rotational and Parker-Rayleigh-Taylor in accretion flows \citep{Kadowaki_2018}.  
Additionally, recent 3D MHD simulations of solar flare current sheets  \citep{Wang_2023}, reported similar reconnection rates, where turbulence  also appears  to be  driven by Kelvin-Helmholtz instability.

We have also analyzed the dependence of $V_\text{rec}$ on the plasma parameter $\beta$. We have also found a weak, albeit noticeable, dependence of the reconnection rate on 
$\beta$, expressed as $\langle V_\text{rec}/V_A \rangle = 0.0372 - 0.0016 \log \beta$, for $\beta$ values ranging from 2 to 64. This dependence closely resembles that obtained by \cite{kowal2017statistics}. 
This correlation was not initially considered in the original theory proposed by \cite{lazarian1999reconnection}. We believe that such dependence arises because lower plasma parameter $\beta$ values in our simulations correspond to lower plasma sound speeds,
resulting in a higher Mach number. This, in turn, makes turbulence more dominant, intensifying the wandering of magnetic field lines and enhancing the turbulent reconnection rate.


We have also investigated the relationship between the reconnection rate and the magnetic Prandtl number, but did not find any clear dependency, once turbulence is fully developed in the domain,  for   values of $\text{Pr}_m$ ranging from 1 to 60. This is in contrast to earlier predictions based on 2D MHD simulations \citep{park1984reconnection, Comisso_2015}, but align with recent 3D MHD simulations by \citet{Brandenburg2024} and theoretical predictions of \cite{Jafari_2018} and the original theory of \cite{lazarian1999reconnection}. 

For all the models simulated, with different values of plasma$-\beta$ and magnetic Prandtl numbers, we observe a consistent increase in the reconnection rate over time, reaching quasi-stationary values that align with the same time where the power spectrum coincides with the Kolmogorov turbulent power spectrum. 
This can be clearly observed in the power spectra of simulations with high magnetic Prandtl numbers (Fig. \ref{fig:power_spec_vort_Prm50}).  
In simulations with $\text{Pr}_m = 1$, turbulence develops almost immediately after the injection of the external perturbation due to the rapid growth of the instability. In contrast, in simulations with higher $\text{Pr}_m$, such as in Fig. \ref{fig:power_spec_vort_Prm50}, we can track the gradual growth of the instability following the initial perturbation until it reaches saturation, marking the full development of turbulence.  
Figure \ref{fig:power_spec_vort_Prm50} distinctly shows the transition from the initial non-turbulent stage, characterized by a low reconnection rate, to the Kolmogorov turbulent regime, at the same time  that the reconnection rate reaches its peak (Figure \ref{fig:vrec_plasmoid_turb}). Turbulence remains self-sustained throughout the simulations, persisting well beyond the initial perturbation injection at \( t = 0.1 \, t_A \). It is the key driver of the measured fast reconnection.

Finally, it is also important to note that even with simulations with high Lundquist numbers of $S \sim 10^5$, where tearing instability may develop, we do not observe the formation, merging, and growth of the magnetic islands into huge plasmoids as typically seen in 2D MHD simulations \citep[e.g.,][]{bhattacharjee2009fast, huang2010scaling}. Fig. \ref{fig:jmag_2D_3D} shows 
a slice of the 3D MHD simulation along the $xy-$plane at $t = 2.9 \, t_A$. 
The observed feature could
indeed resemble a plasmoid-like structure in a 2D projection (as seen in 
previous works). However,  along the $xy-$plane, we only observe 
turbulent eddies. Their elongated structure along the z-direction indicates that these eddies are actually a projection of interacting flux tubes within a fully turbulent system
(see  Figs. \ref{fig:2Dcuts_currdens_S1e5_bt2_P0.5_k128} and \ref{fig:2Dcuts_vorticity_S1e5_bt2_P0.5_k128}). Furthermore, this structure
dissipates rapidly rather than growing, as it would be expected in 2D tearing-mode (plasmoid) unstable models.

\section{Conclusions} \label{sec:conclusions}

{Our 3D MHD simulations have established several key findings regarding the dynamics of magnetic reconnection in turbulent environments:}

\begin{itemize}
    \item[1.] Turbulence in the 3D domain, once initiated, is self-sustained, as evidenced by the power spectra slopes consistent with Kolmogorov's turbulence theory, and maintains the reconnection process at enhanced rates.
    
    \item[2.] The reconnection rate reaches a saturation value once turbulence fully develops, reaching a steady state in the 3D domain, for all models simulated. 
    In the models with higher $\text{Pr}_m$, both turbulence and the reconnection rate require more time to develop and  reach steady-state values, as expected, owing to the rise in the viscosity.

    \item[3.] For the simulations with initial forcing, $\text{Pr}_m=1$, and $S=10^5$ we observed an inverse dependence of the reconnection rate on the plasma parameter $\beta$, or equivalently, on the plasma sound speed ($c_s$), which makes turbulence more dominant for the cases with lower $c_s$ and higher Mach numbers, increasing the reconnection rates.

    \item[4.] The reconnection rates obtained in our simulations with initially multi-mode, small-scale perturbations with high Lundquist numbers ($S=10^5$) are of the same order as those obtained in self-generated turbulent models with $S\approx 3 \times 10^3$ \citep{kowal2017statistics}, demonstrating the independence of the reconnection rate on the Lundquist number, as proposed by \cite{lazarian1999reconnection};

    \item[5.] The reconnection rates obtained in our simulated models with self-sustained turbulence are systematically larger (by a factor of 5 or 6) and grow faster than the ones generated solely by resistive tearing mode 
    instability measured from 2D and 3D MHD simulations \citep[e.g.,][]{bhattacharjee2009fast, huang2010scaling, Huang_2016}.

    \item[6.] There is no clear dependency between the reconnection rate and the magnetic Prandtl number, tested for values  ranging from $\text{Pr}_m =1$ to 60. This is in contrast to earlier predictions based on 2D PIC and 2D MHD simulations \citep{Comisso_2015}, but align with recent 3D MHD simulations by \citet{Brandenburg2024} and theoretical predictions of \cite{Jafari_2018} and the original theory of \cite{lazarian1999reconnection}. 


    \item[7.] In our 3D MHD simulations with initial small-scale perturbations, we do not observe the formation, merging, and growth of the magnetic islands into huge plasmoids when slicing the domain perpendicular to the current sheet, which is usually seen in 2D MHD simulations \citep[e.g.,][]{bhattacharjee2009fast, huang2010scaling}. Instead, we notice  the formation of turbulent eddies, 
    to which we attribute the enhancement of the reconnection rate.
    \end{itemize}

These insights not only advance our understanding of magnetic reconnection mechanisms, but also underscore the potential for applying these findings to more accurately predict and model astrophysical phenomena where such processes are critical, like in  plasma heating and particle acceleration.
In particular, they reinforce 
the recent findings obtained from 3D MHD simulations of particle acceleration via turbulence-induced
magnetic reconnection in different magnetized environments \citep{Kowal_2012, delValle_2016, Beresnyak_Li_2016, Kadowaki_2021, Medina-Torrejón_2021, Medina-Torrejón_2023, dalpino2024}. \
Future studies may focus on exploring variable injection scales of turbulence and their effects on the reconnection process to further refine our models.

\begin{acknowledgments}

GHV acknowledges fruitful discussions with Lorenzo Sironi, Anatoly Spitkovsky, and James Stone.  The authors also thank the referee for their valuable comments and for conducting a thorough review of the manuscript. GHV and EMdGDP acknowledge support from the Brazilian Funding Agency FAPESP (grants 2013/10559-5, 2020/11891-7, 2021/02120-0, and 2023/10590-1), EMdGDP also acknowledges support from CNPq (grant 308643/2017-8). GK acknowledges support from FAPESP (grants 2013/10559-5, 2021/02120-0, 2021/06502-4, and 2022/03972-2). AL acknowledges NSF grant AST 2307840. EMdGDP also acknowledges partial support by grant no. NSF PHY-2309135 to the Kavli Institute for Theoretical Physics (KITP) and the fruitful discussions during her stay there. 
The simulations presented in this work were performed using the clusters of the Group of Plasmas and High-Energy Astrophysics at IAG-USP (GAPAE), the Group of Theoretical Astrophysics at EACH-USP (Hydra), and the ACCESS allocation PHY230191: Turbulent Magnetic Reconnection in Astrophysical Flows. The clusters GAPAE and Hydra HPC were acquired with support from FAPESP (grants 2013/10559-5 and 2013/04073-2, respectively).
\end{acknowledgments}

\newpage

\appendix

\section{Reconnection Rate Measurement in 3D Simulations} \label{appendix:rec_rate}

The reconnection rate in three-dimensional simulations can be calculated in  several ways \citep[see, e.g.,][]{kowal2009numerical, Daughton_2014, Huang_2016, Beresnyak_2017, beg2022evolution}. In this work, we adapted the method described by \cite{kowal2009numerical} to calculate the reconnection rate from the unsigned magnetic flux. We integrate $|B_x|$ over the entire plane $x=0$, perpendicular to the current sheet. Since the signed flux through this plane is zero, dividing the unsigned integral by two gives the flux contribution from each polarity. As time evolves, the two initial flux tubes merge, and the amplitude of the reconnecting field, $|B_x|$, decreases due to the reconnection process. Then, the reconnection rate is given by the time derivative of the unsigned magnetic flux:

\begin{eqnarray}
    V_{\text{rec}} =-  \frac{1}{2 |B_{x,0}| L_z}   \partial_t \iint_S |B_x|dA, \label{eq:rec_rate_line_flux}
\end{eqnarray}

\noindent where $|B_{x,0}|$ represents initial amplitude of the non-reconnecting field, which in this configuration is given by $|B_{x,0}| = \max (|B_x|_{x=0}) \sim 1$.

All models tested in this work have $L_z = 0.5$, leading to the following formula for the reconnection rate:

\begin{equation}
    V_\text{rec} = - \dfrac{\partial}{\partial t} \iint_S |B_x|_{x=0} \, dy \, dz, \label{eq:vrec_plane_x=0}
\end{equation}

\noindent
where $S$ is the plane $x = 0$, corresponding to $(y,z) \in [-0.5, 0.5] \times  [-0.25, 0.25]$. Since the reconnection rate in this method is calculated across the plane $x=0$, we should have less numerical effects due to the boundary conditions and the accumulation of magnetic flux at $x = \pm 0.5$.


For comparison,  we can also compute the rate using the method described in \cite{Huang_2016}, where the magnetic flux is calculated along the current sheet mid-plane ($y=0$) as


\begin{equation}
    \Phi_{HB} = \max_{x \in [-0.5, 0.5]}  \int_{-0.5}^x \int_{-0.25}^{0.25} B_y |_{y= 0} \, dz \, dx' , \label{eq:flux_HB16}
\end{equation}

\noindent
and the reconnection rate is simply $V_\text{rec} = \partial_t \Phi_{HB}$.

\begin{figure}[H]
    \centering
    \includegraphics[width = 0.48 \textwidth]{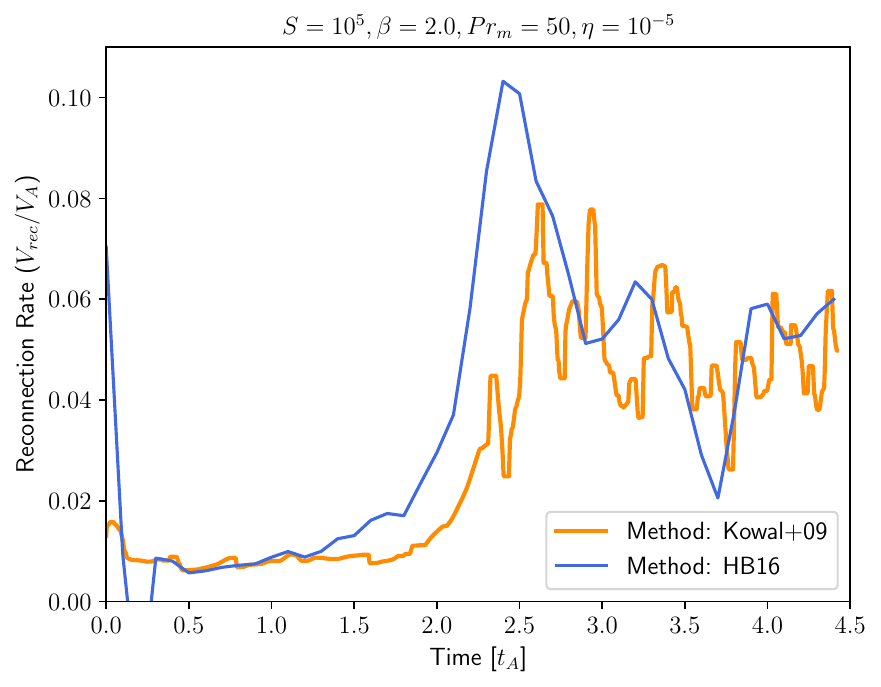}
    \caption{Comparison of the reconnection rate measured with the method described in this paper and in \cite{kowal2009numerical} (orange) and using the method of \cite{Huang_2016} (blue).}
    \label{fig:compar_vrec_BH16}
\end{figure}

In Fig. \ref{fig:compar_vrec_BH16}, we show the time evolution of the reconnection rate for one simulation (with $S = 10^5, \beta = 2.0$ and $\text{Pr}_m=50$) calculated using the two methods described above. The blue line corresponds to the method of \cite{Huang_2016} using Eq. (\ref{eq:flux_HB16}), while the orange line corresponds to the reconnection rate as calculated in this work and initially proposed by \cite{kowal2009numerical}, using Eq. (\ref{eq:vrec_plane_x=0}).

We notice, from Fig. \ref{fig:compar_vrec_BH16}, a good agreement of the reconnection rate measured using both methods. {As in Figure \ref{fig:vrec_plasmoid_turb},  there is an} initial average rate of $\langle V_\text{rec}/V_A \rangle = 0.008(1)$. 
During the formation of a central flux rope between $t=1.5$ and $t=2.5$ in the simulation, the method from \cite{Huang_2016} identifies the magnetic flux within the newly formed flux rope as reconnected. In contrast, the approach used in this work    
detects reconnection only when the flux rope moves away from the $yz-$plane. As a result, the blue curve rises before the orange one.
After $t \sim 2.5$ both methods agree very well. 

\section{Different initial conditions}
\label{appendix:dif_initial_cond}

As discussed in Section \ref{sec:numerics}, we adopted, for our initial configuration, a constant guide field $B_z$ and  non-uniform density in order to maintain the pressure balance. This setup is slightly different from the initial configuration implemented by \cite{Huang_2016, beg2022evolution}, where the initial density is constant and the guide field $B_z$ is non-uniform such that the system is approximately force-balanced. In their case, the initial guide field can be written as

\begin{equation}
    B_z (x,y) = \sqrt{5 \pi^2 \psi^2 + B_{z, \text{min}}^2}, \label{eq:Bz_HB16}
\end{equation}

\noindent
where $\psi$ is given by Eq. (\ref{eq:psi_HB}) and $B_{z, \text{min}}$ is a free parameter.

\begin{figure}[H]
    \centering
    \includegraphics[width = 0.48 \textwidth]{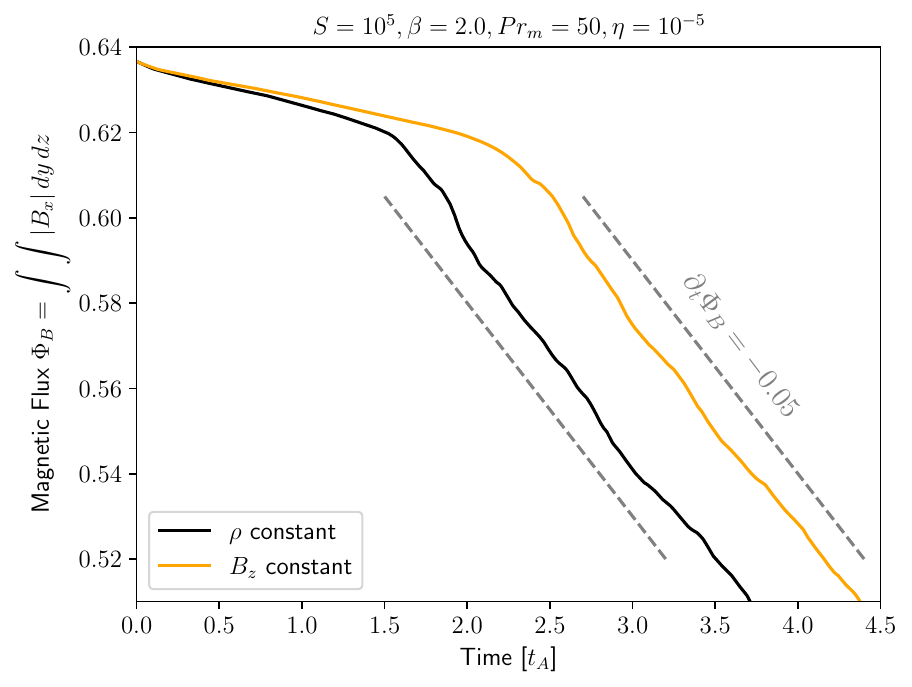}
    \includegraphics[width = 0.47 \textwidth]{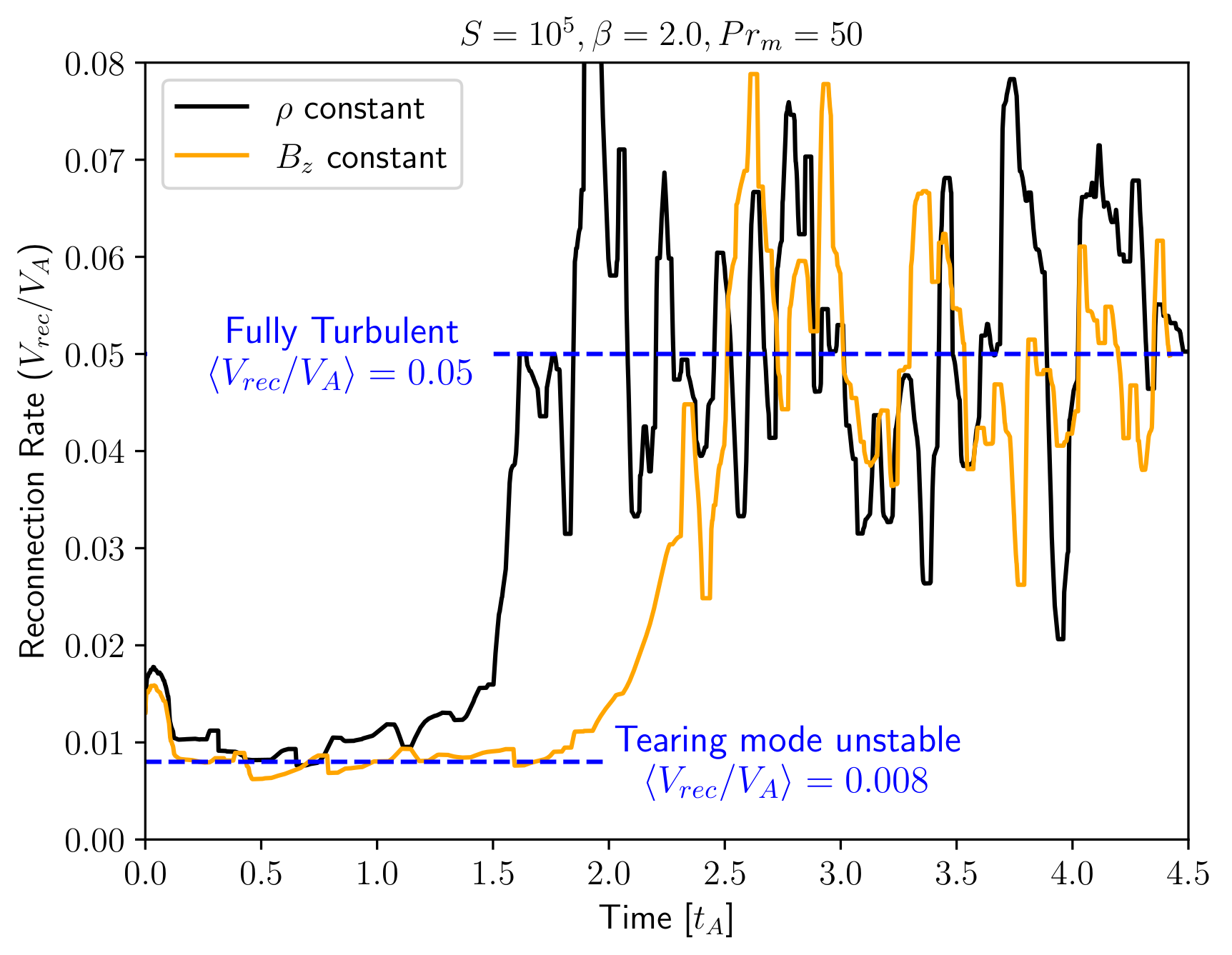}
    \caption{Time evolution of the magnetic flux (left) and the reconnection rate (right) for two different initial conditions, one with constant density (black) and the other with constant guide field, which is used in all simulated models of this work (orange). The physical parameters of both simulations are $S =10^5 $, $\beta = 2.0$, $\text{Pr}_m=50$, $P_\text{inj} = 0.1$, and $k_\text{inj}=128$.}
    \label{fig:dif_initial_cond_BZ_dens_const}
\end{figure}

In Fig. \ref{fig:dif_initial_cond_BZ_dens_const}, we show the time evolution of the magnetic flux (left) and the reconnection rate (right) for two simulations, both with $S = 10^5$, $\beta = 2.0$, and $\text{Pr}_m = 50$. For the simulation represented by the black solid line the initial configuration was the same implemented by \cite{Huang_2016}, where the density is constant and the guide field varies according to Eq. (\ref{eq:Bz_HB16}), with $B_{z, \text{min}} = 0.5$. The simulation represented by the orange solid line, has the setup employed in this work, with the guide field set to be $B_z = 0.5$ (constant), and the density profile {as in  Fig.} \ref{fig:initial-config}. This is the same simulation presented in Fig. \ref{fig:vrec_plasmoid_turb}.

We can notice from Fig. \ref{fig:dif_initial_cond_BZ_dens_const} that, for the simulation with initial constant density (black), turbulence develops faster than in the case of constant guide-field (orange). 
This could be attributed to the fact that an initial constant guide field \( B_z \), rather than a constant density, introduces greater inertia to maintain pressure balance. As a result, the development of turbulence is delayed.
On the other hand, we can observe in both cases the same trend, with the same average reconnection rates: the initial value of $\langle V_\text{rec}/V_A\rangle = 0.008(2)$ due to the development of the tearing/plasmoid instability, and the enhancement of the reconnection rate to an average value of $\langle V_\text{rec}/V_A\rangle = 0.05(1)$ during the fully turbulent regime.

\section{Convergence Analysis}
\label{appendix:convergence_pinj}

We adopted $P_\text{inj} = 0.1$ as the standard injection power of our initial perturbation. In this Appendix, we show a brief analysis of convergence, evidencing the independence of the evolution of the 3D system on the choice of $P_\text{inj}$.

\begin{figure}[h!]
    \centering
    \includegraphics[width = 0.47 \textwidth]{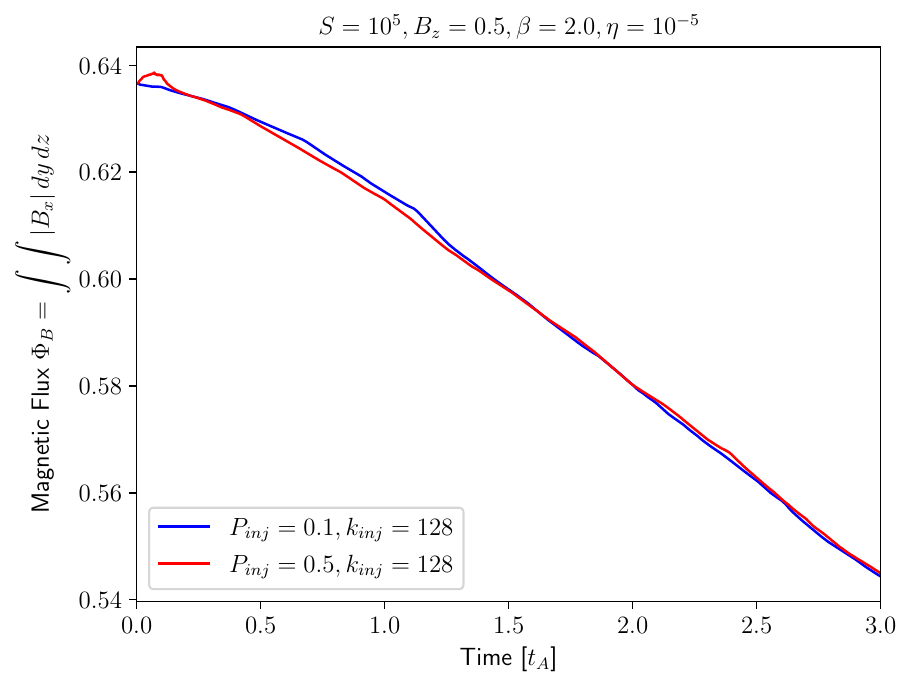} \includegraphics[width = 0.45 \textwidth]{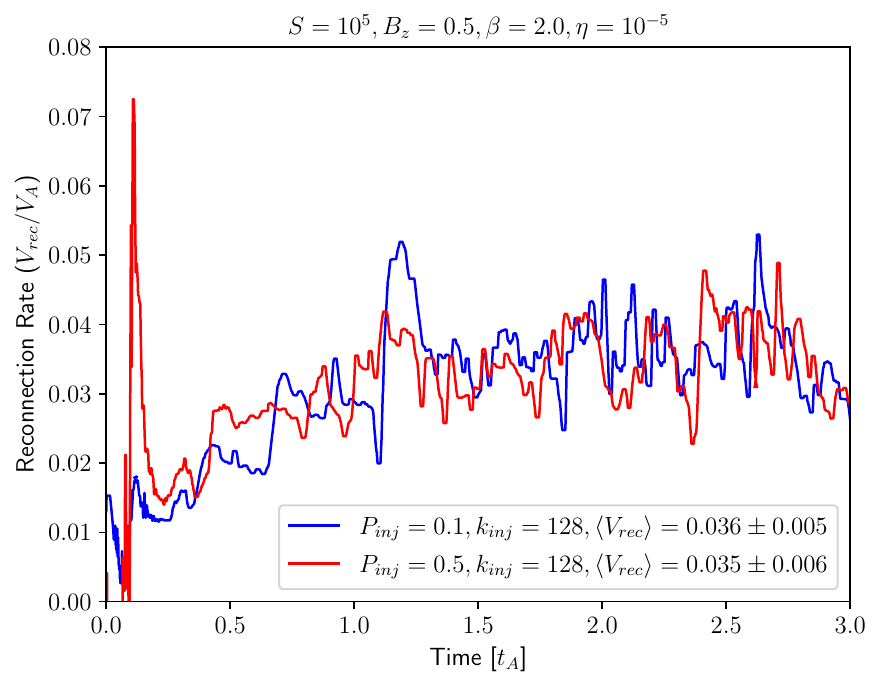}
    \caption{Time evolution of the magnetic flux (left) and the reconnection rate (right) for two different injection powers of the initial perturbation, of $P_{\text{inj}} = 0.1$ (blue) and $0.5$ (red). The other parameters of the simulation are $S =10^5 $, $\beta = 2.0$, $\text{Pr}_m=1$ and $k_\text{inj}=128$.}
    \label{fig:convergence_pinj}
\end{figure}

Figure \ref{fig:convergence_pinj} shows the time evolution of the magnetic flux (left) and the reconnection rate (right) for two simulations, both  with $S = 10^5$, $\beta = 2.0$, $B_z = 0.5$ and $\text{Pr}_m = 1$. The wavenumber of the perturbation was set to $k_\text{inj} = 128$ for both simulations since we desire small-scale perturbation. The difference between the simulations lies in the power of $P_\text{inj} = 0.1$ (blue) and $P_\text{inj} = 0.5$ (red) of the injection of perturbation into the system, from $t = 0$ up to $t = 0.1 \, t_A$.

Despite the initial discrepancy due to the continuous injection up to $0.1 \, t_A$, we notice a good agreement between the two curves of magnetic flux (Fig. \ref{fig:convergence_pinj}, left) for both injection powers, which is reflected in the measurement of the reconnection rate (Fig. \ref{fig:convergence_pinj}, right) of $\langle V_\text{rec} \rangle /V_A = 0.036(5)$ and $0.035(6)$ for $P_\text{inj} = 0.1$ and $0.5$, respectively. Rates are averaged from times between $2.0$ and $3.0 \, t_A$.

\begin{figure}[h!]
    \centering
    \includegraphics[width=0.49\linewidth]{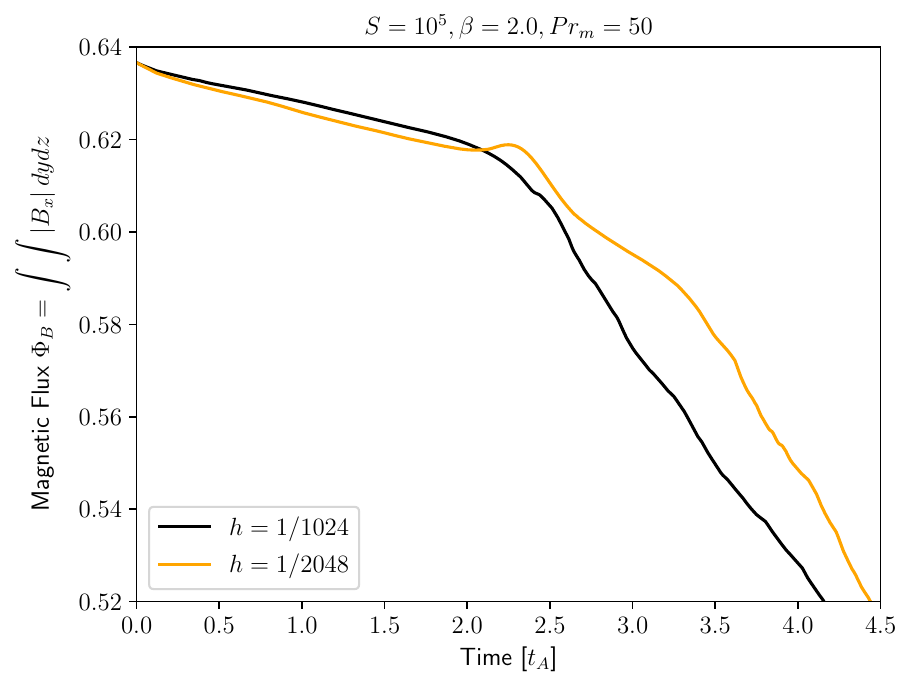}
    \includegraphics[width=0.47\linewidth]{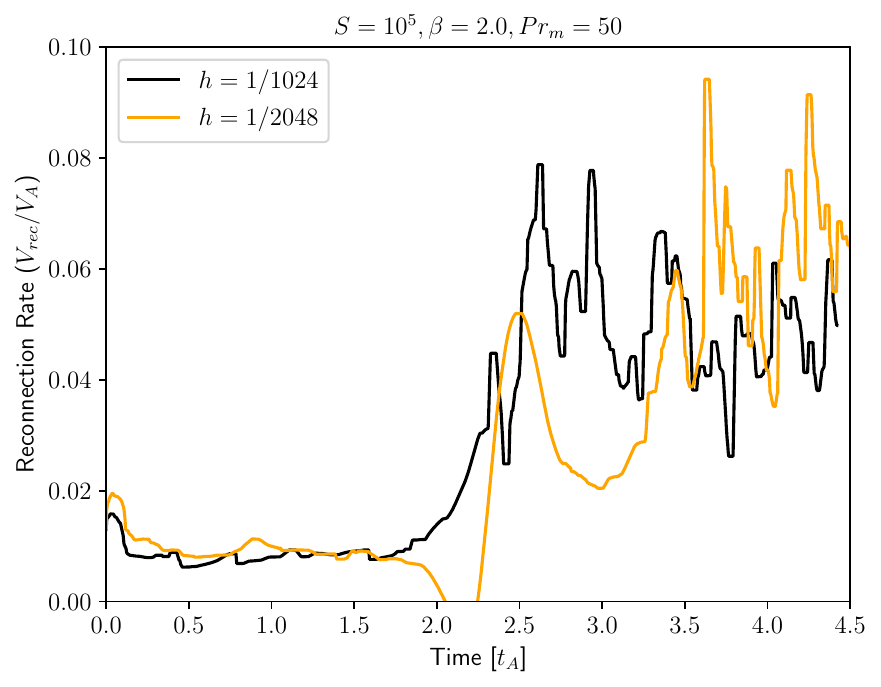}
    \caption{Time evolution of the magnetic flux (left) and the reconnection rate (right) for two different resolutions, with grid size $h = 1/1024$ (black) and $h=1/2048$ (orange). The initial parameters of both simulations are $S =10^5 $, $\beta = 2.0$, $\text{Pr}_m=50$, $B_z = 0.5$ (constant), $P_\text{inj} = 0.1$ and $k_\text{inj}=128$.}
    \label{fig:compar_resolutions_1k_2k}
\end{figure}

We also carried out a convergence analysis considering a simulation with two different resolutions. In Fig. \ref{fig:compar_resolutions_1k_2k}, we show the time evolution of the magnetic flux (left) and the reconnection rate (right) for  grid sizes of $h = 1/1024$ (black) and $h = 1/2048$ (orange). Both simulations were initialized with the same  set of parameters:  $S =10^5 $, $\beta = 2.0$, $\text{Pr}_m=50$, $P_\text{inj} = 0.1$ and $k_\text{inj}=128$, and a constant guide field $B_z = 0.5$.


Despite the minor differences in the temporal evolution of the magnetic flux and the corresponding reconnection rates—an expected outcome given the chaotic nature of the 3D system—we observe a consistent trend in this simulation with a high Prandtl number across both resolutions. Specifically, there is an initial slow reconnection phase up to $t \simeq 2.0$, followed by a significant increase in the reconnection rate as the current layer transitions into the turbulent state.

\bibliography{sample631}{}

\begin{thebibliography}{}
\expandafter\ifx\csname natexlab\endcsname\relax\def\natexlab#1{#1}\fi
\providecommand{\url}[1]{\href{#1}{#1}}
\providecommand{\dodoi}[1]{doi:~\href{http://doi.org/#1}{\nolinkurl{#1}}}
\providecommand{\doeprint}[1]{\href{http://ascl.net/#1}{\nolinkurl{http://ascl.net/#1}}}
\providecommand{\doarXiv}[1]{\href{https://arxiv.org/abs/#1}{\nolinkurl{https://arxiv.org/abs/#1}}}

\bibitem[{Alvelius(1999)}]{alvelius1999random}
Alvelius, K. 1999, Physics of Fluids, 11, 1880

\bibitem[{{Armstrong} {et~al.}(1995){Armstrong}, {Rickett}, \& {Spangler}}]{Armstrong1995ApJ...443..209A}
{Armstrong}, J.~W., {Rickett}, B.~J., \& {Spangler}, S.~R. 1995, \apj, 443, 209, \dodoi{10.1086/175515}

\bibitem[{Balbus \& Hawley(1991)}]{balbus1991powerful}
Balbus, S.~A., \& Hawley, J.~F. 1991, Astrophysical Journal, 376, 214

\bibitem[{Balbus \& Hawley(1998)}]{balbus1998instability}
---. 1998, Reviews of modern physics, 70, 1

\bibitem[{Beg {et~al.}(2022)Beg, Russell, \& Hornig}]{beg2022evolution}
Beg, R., Russell, A.~J., \& Hornig, G. 2022, The Astrophysical Journal, 940, 94

\bibitem[{Begelman(1998)}]{Begelman_1998}
Begelman, M.~C. 1998, The Astrophysical Journal, 493, 291, \dodoi{10.1086/305119}

\bibitem[{Beresnyak(2016)}]{Beresnyak_2017}
Beresnyak, A. 2016, The Astrophysical Journal, 834, 47, \dodoi{10.3847/1538-4357/834/1/47}

\bibitem[{Beresnyak \& Li(2016)}]{Beresnyak_Li_2016}
Beresnyak, A., \& Li, H. 2016, The Astrophysical Journal, 819, 90, \dodoi{10.3847/0004-637X/819/2/90}

\bibitem[{Bhattacharjee {et~al.}(2009)Bhattacharjee, Huang, Yang, \& Rogers}]{bhattacharjee2009fast}
Bhattacharjee, A., Huang, Y.-M., Yang, H., \& Rogers, B. 2009, Physics of Plasmas, 16, 112102

\bibitem[{Birn(1980)}]{birn1980computer}
Birn, J. 1980, Journal of Geophysical Research: Space Physics, 85, 1214

\bibitem[{Biskamp(1986)}]{biskamp1986magnetic}
Biskamp, D. 1986, The Physics of fluids, 29, 1520

\bibitem[{Biskamp(1996)}]{biskamp1996magnetic}
---. 1996, Astrophysics and Space Science, 242, 165

\bibitem[{{Brandenburg} {et~al.}(2024){Brandenburg}, {Neronov}, \& {Vazza}}]{Brandenburg2024}
{Brandenburg}, A., {Neronov}, A., \& {Vazza}, F. 2024, arXiv e-prints, arXiv:2401.08569, \dodoi{10.48550/arXiv.2401.08569}

\bibitem[{Chandrasekhar(1961)}]{chandrasekhar1968hydrodynamic}
Chandrasekhar, S. 1961, Hydrodynamic and hydromagnetic stability (Clarendon Press)

\bibitem[{Chepurnov \& Lazarian(2010)}]{Chepurnov_2010}
Chepurnov, A., \& Lazarian, A. 2010, The Astrophysical Journal, 710, 853, \dodoi{10.1088/0004-637X/710/1/853}

\bibitem[{Comisso {et~al.}(2015)Comisso, Grasso, \& Waelbroeck}]{Comisso_2015}
Comisso, L., Grasso, D., \& Waelbroeck, F.~L. 2015, Physics of Plasmas, 22, 042109, \dodoi{10.1063/1.4918331}

\bibitem[{{Daughton} {et~al.}(2014){Daughton}, {Nakamura}, {Karimabadi}, {Roytershteyn}, \& {Loring}}]{Daughton_2014}
{Daughton}, W., {Nakamura}, T.~K.~M., {Karimabadi}, H., {Roytershteyn}, V., \& {Loring}, B. 2014, Physics of Plasmas, 21, 052307, \dodoi{10.1063/1.4875730}

\bibitem[{Daughton {et~al.}(2009)Daughton, Roytershteyn, Albright, Karimabadi, Yin, \& Bowers}]{Daughton_2009_transition}
Daughton, W., Roytershteyn, V., Albright, B.~J., {et~al.} 2009, Phys. Rev. Lett., 103, 065004, \dodoi{10.1103/PhysRevLett.103.065004}

\bibitem[{Daughton {et~al.}(2011)Daughton, Roytershteyn, Karimabadi, Yin, Albright, Bergen, \& Bowers}]{daughton2011role}
Daughton, W., Roytershteyn, V., Karimabadi, H., {et~al.} 2011, Nature Physics, 7, 539, \dodoi{10.1038/nphys1965}

\bibitem[{{de Gouveia Dal Pino} \& {Medina-Torrejon}(2024)}]{dalpino2024}
{de Gouveia Dal Pino}, E.~M., \& {Medina-Torrejon}, T.~E. 2024, arXiv e-prints, arXiv:2410.13071, \dodoi{10.48550/arXiv.2410.13071}

\bibitem[{{del Valle} {et~al.}(2016){del Valle}, {de Gouveia Dal Pino}, \& {Kowal}}]{delValle_2016}
{del Valle}, M.~V., {de Gouveia Dal Pino}, E.~M., \& {Kowal}, G. 2016, \mnras, 463, 4331, \dodoi{10.1093/mnras/stw2276}

\bibitem[{Dungey(1961)}]{dungey1961interplanetary}
Dungey, J.~W. 1961, Physical Review Letters, 6, 47

\bibitem[{Eyink {et~al.}(2011)Eyink, Lazarian, \& Vishniac}]{Eyink_2011}
Eyink, G.~L., Lazarian, A., \& Vishniac, E.~T. 2011, The Astrophysical Journal, 743, 51, \dodoi{10.1088/0004-637X/743/1/51}

\bibitem[{Forbes(1991)}]{forbes1991magnetic}
Forbes, T. 1991, Geophysical \& Astrophysical Fluid Dynamics, 62, 15

\bibitem[{Furth {et~al.}(1963)Furth, Killeen, \& Rosenbluth}]{furth1963finite}
Furth, H.~P., Killeen, J., \& Rosenbluth, M.~N. 1963, The physics of Fluids, 6, 459

\bibitem[{Gold \& Hoyle(1960)}]{gold1960origin}
Gold, T., \& Hoyle, F. 1960, Monthly Notices of the Royal Astronomical Society, 120, 89

\bibitem[{{Goldreich} \& {Sridhar}(1995)}]{GS_1995}
{Goldreich}, P., \& {Sridhar}, S. 1995, \apj, 438, 763, \dodoi{10.1086/175121}

\bibitem[{Harris(1978)}]{HarrisKaiserBessel1978}
Harris, F. 1978, Proceedings of the IEEE, 66, 51, \dodoi{10.1109/PROC.1978.10837}

\bibitem[{Heyvaerts \& Priest(1984)}]{heyvaerts1984coronal}
Heyvaerts, J., \& Priest, E. 1984, Astronomy and Astrophysics, 137, 63

\bibitem[{Huang \& Bhattacharjee(2010)}]{huang2010scaling}
Huang, Y.-M., \& Bhattacharjee, A. 2010, Physics of Plasmas, 17.
\newblock \url{https://doi.org/10.1063/1.3420208}

\bibitem[{Huang \& Bhattacharjee(2016)}]{Huang_2016}
---. 2016, The Astrophysical Journal, 818, 20, \dodoi{10.3847/0004-637X/818/1/20}

\bibitem[{{Jafari} {et~al.}(2018){Jafari}, {Vishniac}, {Kowal}, \& {Lazarian}}]{Jafari_2018}
{Jafari}, A., {Vishniac}, E.~T., {Kowal}, G., \& {Lazarian}, A. 2018, \apj, 860, 52, \dodoi{10.3847/1538-4357/aac517}

\bibitem[{Kadowaki {et~al.}(2021)Kadowaki, de~Gouveia Dal~Pino, Medina-Torrejón, Mizuno, \& Kushwaha}]{Kadowaki_2021}
Kadowaki, L. H.~S., de~Gouveia Dal~Pino, E.~M., Medina-Torrejón, T.~E., Mizuno, Y., \& Kushwaha, P. 2021, The Astrophysical Journal, 912, 109, \dodoi{10.3847/1538-4357/abee7a}

\bibitem[{Kadowaki {et~al.}(2018)Kadowaki, de~Gouveia Dal~Pino, \& Stone}]{Kadowaki_2018}
Kadowaki, L. H.~S., de~Gouveia Dal~Pino, E.~M., \& Stone, J.~M. 2018, The Astrophysical Journal, 864, 52, \dodoi{10.3847/1538-4357/aad4ff}

\bibitem[{{Kowal} {et~al.}(2011){Kowal}, {de Gouveia Dal Pino}, \& {Lazarian}}]{Kowal_2011}
{Kowal}, G., {de Gouveia Dal Pino}, E.~M., \& {Lazarian}, A. 2011, \apj, 735, 102, \dodoi{10.1088/0004-637X/735/2/102}

\bibitem[{{Kowal} {et~al.}(2012{\natexlab{a}}){Kowal}, {de Gouveia Dal Pino}, \& {Lazarian}}]{Kowal_2012}
---. 2012{\natexlab{a}}, \prl, 108, 241102, \dodoi{10.1103/PhysRevLett.108.241102}

\bibitem[{{Kowal} {et~al.}(2017){Kowal}, {Falceta-Gon{\c{c}}alves}, {Lazarian}, \& {Vishniac}}]{kowal2017statistics}
{Kowal}, G., {Falceta-Gon{\c{c}}alves}, D.~A., {Lazarian}, A., \& {Vishniac}, E.~T. 2017, \apj, 838, 91, \dodoi{10.3847/1538-4357/aa6001}

\bibitem[{{Kowal} {et~al.}(2020){Kowal}, {Falceta-Gon{\c{c}}alves}, {Lazarian}, \& {Vishniac}}]{Kowal_2020}
---. 2020, \apj, 892, 50, \dodoi{10.3847/1538-4357/ab7a13}

\bibitem[{{Kowal} {et~al.}(2009){Kowal}, {Lazarian}, {Vishniac}, \& {Otmianowska-Mazur}}]{kowal2009numerical}
{Kowal}, G., {Lazarian}, A., {Vishniac}, E.~T., \& {Otmianowska-Mazur}, K. 2009, \apj, 700, 63, \dodoi{10.1088/0004-637X/700/1/63}

\bibitem[{{Kowal} {et~al.}(2012{\natexlab{b}}){Kowal}, {Lazarian}, {Vishniac}, \& {Otmianowska-Mazur}}]{kowal2012visc}
---. 2012{\natexlab{b}}, Nonlinear Processes in Geophysics, 19, 297, \dodoi{10.5194/npg-19-297-2012}

\bibitem[{Kruskal \& Tuck(1958)}]{kruskal1958instability}
Kruskal, M., \& Tuck, J. 1958, Proceedings of the Royal Society of London. Series A. Mathematical and Physical Sciences, 245, 222

\bibitem[{{Kulpa-Dybe{\l}} {et~al.}(2010){Kulpa-Dybe{\l}}, {Kowal}, {Otmianowska-Mazur}, {Lazarian}, \& {Vishniac}}]{Kulpa-Dybel-2010}
{Kulpa-Dybe{\l}}, K., {Kowal}, G., {Otmianowska-Mazur}, K., {Lazarian}, A., \& {Vishniac}, E. 2010, \aap, 514, A26, \dodoi{10.1051/0004-6361/200913218}

\bibitem[{{Lazarian} {et~al.}(2020){Lazarian}, {Eyink}, {Jafari}, {Kowal}, {Li}, {Xu}, \& {Vishniac}}]{Lazarian2020review}
{Lazarian}, A., {Eyink}, G.~L., {Jafari}, A., {et~al.} 2020, Physics of Plasmas, 27, 012305, \dodoi{10.1063/1.5110603}

\bibitem[{{Lazarian} {et~al.}(2019){Lazarian}, {Kowal}, {Xu}, \& {Jafari}}]{Lazarian2019review}
{Lazarian}, A., {Kowal}, G., {Xu}, S., \& {Jafari}, A. 2019, in Journal of Physics Conference Series, Vol. 1332, Journal of Physics Conference Series (IOP), 012009, \dodoi{10.1088/1742-6596/1332/1/012009}

\bibitem[{Lazarian \& Vishniac(1999)}]{lazarian1999reconnection}
Lazarian, A., \& Vishniac, E.~T. 1999, The Astrophysical Journal, 517, 700

\bibitem[{Lee \& Fu(1985)}]{lee1985theory}
Lee, L., \& Fu, Z. 1985, Geophysical Research Letters, 12, 105

\bibitem[{Lee \& Fu(1986)}]{lee1986multiple}
---. 1986, Journal of Geophysical Research: Space Physics, 91, 6807

\bibitem[{Loureiro {et~al.}(2007)Loureiro, Schekochihin, \& Cowley}]{loureiro2007instability}
Loureiro, N., Schekochihin, A., \& Cowley, S. 2007, Physics of Plasmas, 14, 100703

\bibitem[{{Loureiro} {et~al.}(2012){Loureiro}, {Samtaney}, {Schekochihin}, \& {Uzdensky}}]{Loureiro2012}
{Loureiro}, N.~F., {Samtaney}, R., {Schekochihin}, A.~A., \& {Uzdensky}, D.~A. 2012, Physics of Plasmas, 19, 042303, \dodoi{10.1063/1.3703318}

\bibitem[{Loureiro {et~al.}(2009)Loureiro, Uzdensky, Schekochihin, Cowley, \& Yousef}]{Loureiro2009}
Loureiro, N.~F., Uzdensky, D.~A., Schekochihin, A.~A., Cowley, S.~C., \& Yousef, T.~A. 2009, Monthly Notices of the Royal Astronomical Society: Letters, 399, L146, \dodoi{10.1111/j.1745-3933.2009.00742.x}

\bibitem[{Matthaeus \& Lamkin(1985)}]{matthaeus1985rapid}
Matthaeus, W., \& Lamkin, S. 1985, The Physics of fluids, 28, 303

\bibitem[{{Medina-Torrej{\'o}n} {et~al.}(2021){Medina-Torrej{\'o}n}, {de Gouveia Dal Pino}, {Kadowaki}, {Kowal}, {Singh}, \& {Mizuno}}]{Medina-Torrejón_2021}
{Medina-Torrej{\'o}n}, T.~E., {de Gouveia Dal Pino}, E.~M., {Kadowaki}, L. H.~S., {et~al.} 2021, \apj, 908, 193, \dodoi{10.3847/1538-4357/abd6c2}

\bibitem[{{Medina-Torrej{\'o}n} {et~al.}(2023){Medina-Torrej{\'o}n}, {de Gouveia Dal Pino}, \& {Kowal}}]{Medina-Torrejón_2023}
{Medina-Torrej{\'o}n}, T.~E., {de Gouveia Dal Pino}, E.~M., \& {Kowal}, G. 2023, \apj, 952, 168, \dodoi{10.3847/1538-4357/acd699}

\bibitem[{{Mignone}(2007)}]{HLLD}
{Mignone}, A. 2007, Journal of Computational Physics, 225, 1427, \dodoi{10.1016/j.jcp.2007.01.033}

\bibitem[{Oishi {et~al.}(2015)Oishi, Low, Collins, \& Tamura}]{Oishi_2015}
Oishi, J.~S., Low, M.-M.~M., Collins, D.~C., \& Tamura, M. 2015, The Astrophysical Journal Letters, 806, L12, \dodoi{10.1088/2041-8205/806/1/L12}

\bibitem[{Park {et~al.}(1984)Park, Monticello, \& White}]{park1984reconnection}
Park, W., Monticello, D., \& White, R. 1984, The Physics of fluids, 27, 137

\bibitem[{{Parker}(1957)}]{Parker1957}
{Parker}, E.~N. 1957, \jgr, 62, 509, \dodoi{10.1029/JZ062i004p00509}

\bibitem[{Parker(1983)}]{parker1983magnetic}
Parker, E.~N. 1983, The Astrophysical Journal, 264, 642

\bibitem[{Parker(1988)}]{parker1988nanoflares}
---. 1988, Astrophysical Journal, 330, 474

\bibitem[{Paschmann {et~al.}(1979)Paschmann, Sonnerup, Papamastorakis, Sckopke, Haerendel, Bame, Asbridge, Gosling, Russell, \& Elphic}]{paschmann1979plasma}
Paschmann, G., Sonnerup, B.~{\"O}., Papamastorakis, I., {et~al.} 1979, Nature, 282, 243

\bibitem[{Petschek(1964)}]{petschek1964physics}
Petschek, H. 1964, in Proc. of AAS-NASA Symp., Vol. 425, NASA Spec. Pub.

\bibitem[{Piddington(1974)}]{piddington1974alfven}
Piddington, J. 1974, Solar Physics, 38, 465

\bibitem[{{Ranocha} {et~al.}(2022){Ranocha}, {Dalcin}, {Parsani}, \& {Ketcheson}}]{SSPRK324}
{Ranocha}, H., {Dalcin}, L., {Parsani}, M., \& {Ketcheson}, D.~I. 2022, Communications on Applied Mathematics and Computation, 4, 2661, \dodoi{10.1007/s42967-021-00159-w}

\bibitem[{Sakai \& Ohsawa(1988)}]{sakai1988particle}
Sakai, J.-I., \& Ohsawa, Y. 1988, Space science reviews, 46, 113, \dodoi{10.1007/BF00173877}

\bibitem[{Samtaney {et~al.}(2009)Samtaney, Loureiro, Uzdensky, Schekochihin, \& Cowley}]{Samtaney_2009}
Samtaney, R., Loureiro, N.~F., Uzdensky, D.~A., Schekochihin, A.~A., \& Cowley, S.~C. 2009, Phys. Rev. Lett., 103, 105004, \dodoi{10.1103/PhysRevLett.103.105004}

\bibitem[{Shafranov(1956)}]{shafranov1956stability}
Shafranov, V. 1956, Sov. J. At. Energy, 5, 38

\bibitem[{Singh {et~al.}(2016)Singh, Mizuno, \& Dal~Pino}]{singh2016spatial}
Singh, C.~B., Mizuno, Y., \& Dal~Pino, E. M. d.~G. 2016, The Astrophysical Journal, 824, 48

\bibitem[{Speiser(1965)}]{speiser1965particle}
Speiser, T. 1965, Journal of Geophysical Research, 70, 4219

\bibitem[{{Suresh} \& {Huynh}(1997)}]{mp5}
{Suresh}, A., \& {Huynh}, H.~T. 1997, Journal of Computational Physics, 136, 83, \dodoi{10.1006/jcph.1997.5745}

\bibitem[{{Sweet}(1958)}]{Sweet1958}
{Sweet}, P.~A. 1958, in Electromagnetic Phenomena in Cosmical Physics, ed. B.~{Lehnert}, Vol.~6, 123

\bibitem[{Takamoto {et~al.}(2015)Takamoto, Inoue, \& Lazarian}]{takamoto2015turbulent}
Takamoto, M., Inoue, T., \& Lazarian, A. 2015, The Astrophysical Journal, 815, 16

\bibitem[{Takamoto \& Lazarian(2016)}]{Takamoto_2016}
Takamoto, M., \& Lazarian, A. 2016, The Astrophysical Journal Letters, 831, L11, \dodoi{10.3847/2041-8205/831/2/L11}

\bibitem[{Uzdensky \& Kulsrud(2000)}]{uzdensky2000two}
Uzdensky, D., \& Kulsrud, R. 2000, Physics of Plasmas, 7, 4018

\bibitem[{Wang {et~al.}(2023)Wang, Cheng, Ding, Liu, Liu, \& Zhu}]{Wang_2023}
Wang, Y., Cheng, X., Ding, M., {et~al.} 2023, The Astrophysical Journal Letters, 954, L36, \dodoi{10.3847/2041-8213/acf19d}

\end{thebibliography}
\bibliographystyle{aasjournal}



\end{document}